\documentclass[11pt, a4paper]{article}
\usepackage{amssymb,amsmath,mathtools,xcolor,graphicx,xspace,colortbl,ragged2e,rotating}
\usepackage{amsfonts}
\usepackage{setspace}

\usepackage{graphics,color}
\usepackage{makeidx}
\usepackage[english]{babel}
\usepackage{float}
\usepackage{caption}
\usepackage[lofdepth,lotdepth]{subfig}
\usepackage{geometry}
 
\usepackage[authoryear,round]{natbib}
\setcitestyle{aysep={},citesep={;}}
\usepackage{hyperref}              %
\hypersetup{
    bookmarksnumbered=true,
    colorlinks=true,
    linkcolor=blue,
    citecolor=blue,
    filecolor=magenta,      
    urlcolor=cyan,   
    pdfpagemode=FullScreen,        
    }                              %
\DeclareGraphicsExtensions{.pdf,.eps,.ps,.png,.jpg,.jpeg}

\newtheorem{proposition}{Proposition}
\newtheorem{definition}{Definition}
\newtheorem{example}{Example}

\newtheorem{corollary}{Corollary}

\begin{document}

\title{{\Large {Coarse Agents and Intergroup Phenomena}\thanks{I thank audiences of seminars TOM at PSE, Organizational Economics Lunch at MIT, Microeconomics Working Group at EUI, Theory Discussion Group at Bocconi, at Peking University HSBC Business School, at Universidad Carlos III and at JKU Linz, and conferences such as the Transatlantic Theory Workshop 2024, TSE--PSE Workshop 2025, Doctorissimes Conference 2025 (Paris 1), European Meeting in Game Theory 2025 (Maastricht), CEPET (Udine), World Congress of the Econometric Society (Seoul), EEA summer meeting 2025 (Bordeaux), Italian Junior Theory Workshop (CSEF), EAYE (Bilbao), and FUR (Alicante). I am indebted to Philippe Jehiel, Pierpaolo Battigalli, Rachel Kranton, Ran Spiegler, Evan Friedman, Bernard Caillaud, Bob Gibbons, Charles Angelucci, Larry Samuelson, Frederic Koessler, Roberto Corrao, Giacomo Lanzani, Laura Carlotta Hoenig, Aviman Satpathy, Alice Mazzacurati, and Manpreet Singh for useful suggestions and discussions. Funded by the European Union (ERC, TRAITS-GAMES, 101142844). Views and opinions expressed are however those of the author(s) only and do not necessarily reflect those of the European Union or the European Research Council Executive Agency. Neither the European Union nor the granting authority can be held responsible for them. This work has also benefited from the support of the European Union – Next Generation EU, Mission 4 Component 2, CUP J53D23004180006.}}}

\author{{\large {Giacomo Weber}
\thanks{Bocconi University, Via Roentgen 1, 20136 Milan, Italy; giacomo.weber@unibocconi.it }}}

% Confirm the TRAITS-GAMES funding wording before posting.
\maketitle
                              
\vspace{-11mm}
\begin{center}
    \small{Latest version \href{https://drive.google.com/file/d/1qdr-qhQxYFzqbaIIMktdhQp7ZpoZBKbX/view?usp=drive_link}{here} }
\end{center}

%\bigskip
\begin{abstract}
\noindent This paper proposes a framework for analyzing intergroup phenomena based on group-dependent expectations. Heterogeneous players belong to two groups and are sorted into pairs to play a static game. Each player observes her partner and the characteristics that enter her payoff. When matched within her group (ingroup), she holds correct equilibrium expectations about her partner's behavior. When matched across groups (outgroup), she forms coarse expectations. In an Outgroup Homogeneity Equilibrium, these expectations coincide with the aggregate behavior of the outgroup, including members matched ingroup. Who is paired with whom thus shapes the behavior entering these expectations. The framework is illustrated in a coordination game and applied to an organization where the groups are departments and each pair performs a joint task. With departments and bonuses fixed, a designer forms teams to maximize the expected number of successful tasks. Each worker's effort is more productive when her teammate contributes more. Two workers with low effort costs can therefore elicit high effort from one another when paired ingroup. Their contributions raise average effort in their department. Workers in the other department who judge an outgroup teammate by that average may then put more effort into their own tasks.
 
%\noindent \textbf{Keywords}:  \newline
%\textbf{JEL Classification Numbers:} 
\end{abstract}

%--------------------%
%--  INTRODUCTION  --%
%--------------------%
\newpage

%----------------------------------

\section{Introduction}

A person may know whom she faces and still be unsure what to expect from her partner. Economic life brings people together across group boundaries as well as within them. At work these boundaries may separate departments or plants; outside work they may follow neighborhoods or communities. Group membership can matter for what people want to do, through the norms they associate with a social identity \citep{AkerlofKranton2000}. It can also matter for how others see them. A stereotype emphasizes traits that distinguish one group from another \citep{BordaloCoffmanGennaioliShleifer2016}; an employer who cannot observe an applicant's productivity may draw on what she knows about the applicant's group \citep{Phelps1972,Arrow1973}. Here the question concerns a different judgment: even when a person knows whom she faces, how does group membership shape what she expects that partner to do?

Consider an engineer assigned to review a design with a colleague. It is worth studying the drawings more closely if her colleague will come prepared to discuss the details. Within her own department she may know which colleagues usually come well prepared. For another department, the records she sees might give only the average time its engineers spend preparing for a review. She knows her colleague's name and specialty, but turns to that figure to judge how much preparation to expect.

This contrast draws on what social psychologists call the \emph{outgroup homogeneity effect}: people often perceive less variation among members of another group than among members of their own \citep{QuattroneJones1980}. In an experiment by \citet{ParkRothbart1982}, members of three campus sororities saw more differences within their own sorority than members of the other two saw in that same group.\footnote{The tendency varies across settings \citep{MullenHu1989}. Proposed explanations include differences in experience with group members \citep{linville1989perceived} and differences in how people organize what they remember about individuals in each group \citep{OstromEtAl1993differential}.}

What preparation time would the records show? That depends on how the reviews are staffed. Two engineers from the same department who expect one another to prepare thoroughly may both put long hours into their review. Those hours count just as much toward their department's average as hours spent on reviews with other departments. In comparing team assignments, a manager must therefore ask what preparation each would sustain: in equilibrium, the figure used to judge a colleague from another department must equal the average time its engineers choose to put in under that assignment. Which teams should the manager form to help more projects succeed? And when partners decide separately whether to invest in a project, can investment by one pair make it worthwhile for another pair to invest?

To study these questions, consider two fixed groups of players matched in pairs to play a static game. Each observes her partner and the characteristics of both players that matter for her payoff. The partners choose their actions simultaneously. If the partner belongs to the same group (an \emph{ingroup} match), she forms an expectation specific to that individual. If the partner belongs to the other group (an \emph{outgroup} match), she draws on the behavior of her partner's group as a whole. Every member counts equally, including those paired ingroup. In an \emph{Outgroup Homogeneity Equilibrium} (OHE), each player chooses optimally given what she expects her partner to do. Ingroup, she assigns the correct probability to each of her partner's actions. Outgroup, she assigns each action its average probability across the members of her partner's group. That average need not describe the particular partner she faces.

These expectations have a learning interpretation. Suppose each new generation receives a record of how its predecessors played. For an ingroup partner the record describes the choices of those who previously occupied that partner's role. For an outgroup partner it gives only the frequencies of actions across the partner's group. Players use these records to form expectations about how their current partners will act. Suppose there are finitely many actions and small private variations in the payoffs from choosing them. A pattern of play is stationary if the records it generates lead the next generation to choose each action just as often in each role. As the payoff variations vanish, any limit of these stationary patterns is an OHE.\footnote{For a fixed matching in the unperturbed finite-action game, OHE coincides with analogy-based expectation equilibrium \citep{Jehiel2005} for a partition with one class for each other ingroup player and one for the whole outgroup. Self-confirming equilibrium ties the restrictions on a player's conjecture to the feedback she receives \citep{battigalli1987comportamento,FudenbergLevine1993SCE,BattigalliEtAl2016Feedback}. A report might correctly record how often a group invests without revealing how often a particular partner invests. Under OHE, the player uses the group frequency as the probability that her own partner invests.}

When might uncertainty about whom one will meet lead to the same expectations about a partner's behavior? Suppose the two groups are of equal size, everyone is paired across groups, and once both actions and a player's own characteristics are given, her payoff does not depend on her partner's characteristics (the private-values case). Compare a fixed matching in which each player sees her own partner with one drawn uniformly at random from all matchings across the two groups. In the latter case, players are told only that their partner belongs to the other group. A player facing the random draw gives everyone in the other group the same weight because she is equally likely to meet any of them. With an observed partner, the player uses those same weights even though she knows whom she faces. The two settings therefore give exactly the same equilibrium predictions about how each individual plays. Their implications can differ, however, when some partners are matched ingroup or when a partner's characteristics enter payoffs directly.

How should a manager form teams when their success depends on effort and bonuses are fixed? Departments provide a natural setting for this question. In studies of product development, people in different departments interpret the same project in different ways \citep{Dougherty1992}, and ties across units affect how teams find and use knowledge held elsewhere in the firm \citep{Hansen1999}. The model takes these unit boundaries as given. It considers a problem of moral hazard in teams \citep{Holmstrom1982}: workers differ in their cost of effort, each receives a fixed bonus if her team's task succeeds, and her effort is more productive the more her teammate contributes. The manager assigns workers from two departments to two-person tasks to maximize the expected number of successful tasks.\footnote{Team composition is also central to \citet{Prat2002}, who asks whom a principal should hire when agents receive signals that can differ in their correlation. Here the workforce is fixed; changing assignments changes the behavior on which workers in other teams base their expectations.}

Consider pairing two low-cost workers from the same department. They elicit high effort from one another, which helps their own task succeed. That effort also counts toward their department's average. A worker from the other department who meets an outgroup teammate relies on this average, so keeping the two low-cost workers together can encourage effort on tasks to which neither is assigned. Giving each an outgroup teammate instead would make each respond to the other department's average, at the cost of breaking their productive partnership.

A simple rule emerges as the departments grow. Suppose they are equally large, all costs lie in a fixed bounded interval where effort choices are interior, and both cost distributions approach the same distribution with a density. In every optimal matching, a common cutoff describes all but a vanishing fraction of workers: those below it work ingroup and those above it work outgroup. Both kinds of team account for a positive share of tasks in the limit. To see where the cutoff comes from, compare working with an equally costly colleague in one's own department with working across departments. At the cutoff, the effort that the ingroup colleague would exert is just the average effort in the other department. Below it, an equally low-cost colleague elicits more effort; above it, the group average does. In large enough organizations, the designer can thus achieve more expected successes than under the best assignment when everyone correctly anticipates her own teammate's effort. Below the cutoff, the two limiting effort schedules agree. The gain comes from above the cutoff, where working outgroup elicits more effort.

Coordination games pose a familiar question: what can single out one pattern of play when each player's preferred action depends on what she expects her partner to do? In the electronic mail game of \citet{rubinstein1989electronic}, a message about which game is being played may fail to arrive. In the global games of \citet{CarlssonVanDamme1993}, players receive noisy observations of the payoffs. Here each player sees the characteristics that determine the return to investing with her partner. What she needs to judge is whether the partner will invest. All matches are outgroup, but the return to joint investment increases with either partner's characteristic, so the private-values benchmark no longer applies. Consider two groups with the same list of characteristics, with players of the same rank paired across groups. In the example, each member of the highest pair finds it worthwhile to invest even if her partner does not. Although no one else is matched with either of them, their investment raises the probability that each member of the next pair assigns to investment by her own partner. That makes investment worthwhile for both. The argument works down the ranks: investment by everyone is the unique OHE. Other distributions of returns can instead single out investment by only the higher-return pairs.

Could uncertainty about whom one will meet give the same prediction? Suppose the signal reveals the identity of an ingroup partner but reports only the group of an outgroup partner. For the population in the full-investment example, every common prior over matchings admits both a pure Bayesian equilibrium with full investment and another in which a player does not invest after receiving a signal of positive probability. In this benchmark, the weight a player gives a possible partner is the probability of meeting that person conditional on her signal. Under outgroup homogeneity, the investment of the highest pair bears on other players' decisions even though those two face only each other at the fixed matching.

%-----------------%
%--  FRAMEWORK  --%
%-----------------%
\section{Framework}\label{section:Framework}

\subsection{Strategic Environment}\label{section:strategic_environment}

Consider a finite set \(I\) containing an even number of players, who interact simultaneously in disjoint pairs. The pairwise interactions are governed by a complete matching \(\phi\), with \(\phi(i)\) denoting the match of player \(i\). Thus, \(\phi(\phi(i))=i\) and \(\phi(i)\ne i\) for every \(i\in I\). The set of all such matchings is denoted by \(\Phi\). The matching is fixed before players choose their actions.\footnote{Depending on the application, it may be specified exogenously or selected by a designer.}

Each player \(i\) has payoff-relevant individual characteristics \(x_i\in X\) and available actions \(a_i\in A_i\). The action space is assumed to be common to all players: \(A_i=A\), where \(A\) is nonempty and finite. The characteristic and action profiles are denoted by \(x=(x_i)_{i\in I}\in X^I\) and \(a=(a_i)_{i\in I}\in A^I\), respectively.

Player \(i\)’s individual consequence belongs to a space \(Y_i\) and is evaluated by a von Neumann–Morgenstern utility function \(v_i:Y_i\to\mathbb R\). A profile of consequences is a vector \(y=(y_i)_{i\in I}\in Y=\prod_{i\in I}Y_i\), and player \(i\)’s utility depends only on \(y_i\). The outcome function
\[
f:A^I\times X^I\times\Phi\longrightarrow\Delta(Y)
\]
assigns to each action profile, characteristic profile, and matching a lottery over consequences. Deterministic outcomes are included as degenerate lotteries. Pairwise interactions require that the marginal distribution of \(y_i\) depend only on \(a_i,a_{\phi(i)},x_i,x_{\phi(i)}\).\footnote{The joint lottery may exhibit dependence across (matched) players’ consequences: for instance, in one of the applications, matched agents share the success or failure of a team task.}

For a lottery \(P\in\Delta(Y)\), player \(i\)’s expected utility is
\[
V_i(P)=\int_Y v_i(y_i)\,P(dy).
\]
Outcome spaces and utility functions are assumed measurable, and expected utilities are finite.

The matching is part of the rules of interaction: it specifies whose action
and characteristics enter the distribution of each player's consequence.
For given \(x\) and \(\phi\), let \(f^{x,\phi}(a)=f(a,x,\phi)\).
The outcome function \(f^{x,\phi}\), together with the players and their
action and consequence spaces, specifies the static game form. Adding the
players' utility functions gives the static game. The description retains the matching explicitly:
\[
\Gamma^{x,\phi}
=\left\langle I,\Delta(Y),(A_i)_{i\in I},\phi,f^{x,\phi},(V_i)_{i\in I}
\right\rangle.
\]
The payoff of player \(i\) matched with \(j\) is denoted by \(u_i(a_i,a_j;x_i,x_j)\). By the pairwise restriction,
\[
V_i\!\left(f(a,x,\phi)\right)
=
u_i(a_i,a_{\phi(i)};x_i,x_{\phi(i)}).
\]

Let \(\sigma_i\in\Delta(A)\) denote a mixed action of player \(i\), and \(\sigma=(\sigma_i)_{i\in I}\in(\Delta(A))^I\) denote a profile of mixed actions. Under independent action randomizations, the expected payoff of player \(i\) matched with \(j\) is
\[
U_i(\sigma_i,\sigma_j;x_i,x_j)
=
\sum_{a_i,a_j\in A}
\sigma_i(a_i)\sigma_j(a_j)\,
u_i(a_i,a_j;x_i,x_j).
\]

In the baseline model, each player \(i\) observes the identity of her
partner, \(\phi(i)\), and the characteristics \(x_i\) and \(x_{\phi(i)}\)
before choosing an action. She knows the common action set \(A\), her utility function \(v_i\), and how the actions and characteristics of her pair determine the distribution of her individual consequence.\footnote{One may alternatively assume that players observe the entire matching. Under the conjecture restriction introduced below, this leaves the set of equilibrium mixed-action profiles unchanged for each realized matching.}

Given \(x\) and \(\phi\), a mixed-action profile \(\sigma\) is a Nash equilibrium if, for every \(i\in I\) and every \(\sigma_i'\in\Delta(A)\),
\[
U_i(\sigma_i,\sigma_{\phi(i)};x_i,x_{\phi(i)})
\ge
U_i(\sigma_i',\sigma_{\phi(i)};x_i,x_{\phi(i)}).
\]

\subsection{Groups and Outgroup Homogeneity Equilibrium}\label{section:groups_and_OHE}

Group membership shapes what a player expects her partner to do. She distinguishes among partners from her own group, but assesses a partner from the other group on the basis of that group's aggregate behavior.

Formally, let \(G_1,G_2\subseteq I\) be two labelled groups satisfying
\(G_1\cap G_2=\varnothing\) and  \(G_1\cup G_2=I\). Group membership is fixed, and \(G(i)\) denotes the group containing player \(i\). Each player recognizes her own and her match's group membership.

Group membership and matching are distinct: a player may interact with someone from her own group or from the other group. The matching therefore determines whether the interaction is ingroup or outgroup. 

Given \(\phi\), player \(i\) is \emph{matched ingroup} if \(\phi(i)\in G(i)\), and \emph{matched outgroup} otherwise. 

For each nonempty group \(G\), we denote the corresponding sets of players by
\[
\operatorname{In}_G^\phi=\{i\in G:\phi(i)\in G\},
\qquad
\operatorname{Out}_G^\phi=\{i\in G:\phi(i)\notin G\}.
\]

To evaluate her available actions, a player forms a conjecture about how her match will behave. This conjecture is represented by a probability distribution over the common action set. Let \(\beta_i\in\Delta(A)\) denote player \(i\)'s conjecture, and let \(\beta=(\beta_i)_{i\in I}\) be a profile of conjectures.

The behavioral restriction concerns how these conjectures relate to actual behavior. In ingroup matches, conjectures are calibrated to the individual partner. In outgroup matches, they are calibrated to aggregate behavior in the partner's group. To describe this aggregate, given a mixed-action profile \(\sigma\), define for each nonempty group $G$,
\[
\bar\sigma_G=\frac{1}{|G|}\sum_{j\in G}\sigma_j
\in\Delta(A).
\]
This distribution gives equal weight to every member of \(G\), including members who are themselves matched ingroup.

Let \(\beta\) be \emph{consistent} with \(\sigma\) under \(\phi\) if, for every player \(i\),
\[
\beta_i=
\begin{cases}
\sigma_{\phi(i)}, & \text{if }\phi(i)\in G(i),\\[2mm]
\bar\sigma_{G(\phi(i))}, & \text{if }\phi(i)\notin G(i).
\end{cases}
\]

An ingroup conjecture equals the actual partner's mixed action. An outgroup conjecture equals the group average and need not coincide with the actual partner's mixed action. Consistency thus captures a stark version of the outgroup homogeneity effect through the asymmetry in players' perceptions. At a given profile \(\sigma\), outgroup conjectures do not distinguish between different members of the outgroup, even though players observe their partner's identity and characteristics. 

A player's conjecture specifies the probabilities she assigns to her partner's actions. She evaluates her own actions using those probabilities and the actual characteristics \(x_i\) and \(x_{\phi(i)}\).

Given \(\phi\), a mixed action \(\sigma_i\) is a \emph{best response} to \(\beta_i\) if, for every \(\sigma_i'\in\Delta(A)\),
\[
U_i(\sigma_i,\beta_i;x_i,x_{\phi(i)})
\geq
U_i(\sigma_i',\beta_i;x_i,x_{\phi(i)}).
\]
The conjecture is held fixed when evaluating deviations. In other words, player \(i\) chooses optimally as if her actual partner played according to \(\beta_i\).

\begin{definition}[Outgroup Homogeneity Equilibrium.]
    Given the strategic environment described by \(\Gamma^{x,\phi}\) and given groups \((G_1,G_2)\), a mixed-action profile \(\sigma\in(\Delta(A))^I\) is an \emph{Outgroup Homogeneity Equilibrium} (OHE) under \(\phi\) if there exists a conjecture profile \(\beta\in(\Delta(A))^I\) such that: (i) for every \(i\in I\), \(\sigma_i\) is a best response to \(\beta_i\); and (ii) \(\beta\) is consistent with \(\sigma\) under \(\phi\).
\end{definition}

An OHE thus combines optimal choice with asymmetric perception. In ingroup matches, players best respond to their actual match's mixed action. In outgroup matches, they best respond to the group aggregate applied to that match. Their conjectures therefore agree with the group average but can differ from the mixed action of the partner they face.

\begin{proposition}
\label{prop:ohe-existence}
Consider a finite game described by \(\Gamma^{x,\varphi}\) with characteristic profile \(x\in X^I\) and matching
\(\varphi\in\Phi\), and let two disjoint groups \(G_1,G_2\)
cover \(I\). There exists an OHE
\(\sigma\in(\Delta(A))^I\) under \(\varphi\).
\end{proposition}

The result follows from a standard application of Kakutani's
fixed-point theorem to the correspondence of best responses to
consistent conjectures. The proof is in Appendix~\ref{proof:ohe-existence}.

\textbf{Nash equilibrium as a special case.}
If \(G_1=I\) or \(G_2=I\), then, for every matching \(\phi\in\Phi\), the set of OHE profiles coincides with the set of Nash equilibria in mixed actions of the game described by \(\Gamma^{x,\phi}\). Indeed, every player is matched ingroup, so consistency requires \(\beta_i=\sigma_{\phi(i)}\) for every \(i\in I\). The OHE best-response conditions are therefore exactly the Nash equilibrium conditions. 

More generally, the two definitions coincide whenever the matching $\phi$ is such that all agents are paired \textit{ingroup}. Clearly, the equilibrium actions prescribed by OHE can differ from Nash equilibrium for general groups and matchings.

\subsection{Learning foundations}
\label{subsec:learning-foundation_ABEE}

For the finite-action environment, a population learning interpretation
clarifies how conjectures can be informed by past behavior. Fix the roles
$I$, characteristics $x$, groups, matching $\phi$, and baseline payoff
functions. In each period $t\geq 1$, a fresh population of mass one occupies
each role $i$. Each subject interacts once with a subject occupying role
$\phi(i)$, receives her consequence, and leaves. The matching across roles
is fixed throughout. Write $\sigma_i^t\in\Delta(A)$ for the action distribution
in role $i$ in period $t$, and $\bar\sigma_G^t$ for the corresponding group
average. Thus the distributions that describe mixed actions in the static
model describe population action shares here.

Before choosing an action, each subject has access to the preceding period's
action frequencies. She observes the frequencies separately for every other
role in her own group. For the opposite group, when nonempty, she observes
only the frequencies pooled across all its roles, including those matched
ingroup. Since the roles have equal mass and each subject chooses one
action, the pooled distribution is the arithmetic average of the role
distributions. An arbitrary initial profile $\sigma^0$ determines the
frequencies available in the first period; subjects need not observe the
full profile.

Each subject knows her own payoff rule and observes her partner's role,
characteristics, and group membership. She uses the preceding period's
frequencies for her partner's role if matched ingroup and for her partner's
group if matched outgroup. Her conjecture is therefore
\[
\beta_i^t=
\begin{cases}
\sigma_{\phi(i)}^{t-1}, & \phi(i)\in G(i),\\[1mm]
\bar\sigma_{G(\phi(i))}^{t-1}, & \phi(i)\notin G(i).
\end{cases}
\]
The learning model assumes that each subject uses these past frequencies as her conjecture about her current partner's action. The actual characteristics $x_i,x_{\phi(i)}$ remain in the payoff
calculation. Subjects may observe their own actions and realized
consequences after play. Their successors have access only to the specified
action frequencies, not to personal interaction histories.

To obtain population mixtures from individually optimal pure choices,
introduce small private payoff perturbations, following the approach of
\citet{Harsanyi1973} and its use in learning by \citet{FudenbergKreps1993}.
Fix $\varepsilon>0$ throughout a learning process. Before acting, each
subject in role $i$ privately observes a vector
$z_i=(z_i(a))_{a\in A}$ with independent uniform $[0,1]$ coordinates. Her
expected payoff conditional on actions $a_i,a_{\phi(i)}$ is
\[
u_i(a_i,a_{\phi(i)};x_i,x_{\phi(i)})+\varepsilon z_i(a_i).
\]
Here $\varepsilon$ measures the perturbation size, and $z_i(a)$ is the
subject's private payoff shock to action $a$. The vector is subject-specific;
the role subscript does not denote a common shock.
Each cohort has exactly the specified shock distribution, and partners'
private vectors are independent. Subjects choose pure actions maximizing
perturbed expected payoff given $\beta_i^t$. The maximizing action is
unique almost surely, so $\sigma_i^t$ is induced by the distribution of
shocks. Appendix~\ref{appendix:learning-foundation} gives the population and encounter construction.

\begin{definition}[Payoff-perturbed OHE]
\label{def:payoff-perturbed-ohe}
For the specified private payoff shocks and a fixed $\varepsilon>0$, a
profile $\sigma\in(\Delta(A))^I$ is a \emph{payoff-perturbed OHE at scale
$\varepsilon$} under $\phi$ if there exists a conjecture profile $\beta$
that is consistent with $\sigma$ under the original ingroup/outgroup rule
and satisfies, for every $i\in I$ and $a_i\in A$,
\begin{equation}
\label{eq:payoff-perturbed-ohe-choice}
\begin{aligned}
\sigma_i(a_i)
&=\Pr_{\,z_i\sim\operatorname{Unif}([0,1]^A)}
\!\left[
a_i\in\arg\max_{a'_i\in A}
\left\{
\sum_{a_{\phi(i)}\in A}
\beta_i(a_{\phi(i)})\,
u_i(a'_i,a_{\phi(i)};x_i,x_{\phi(i)})
+\varepsilon z_i(a'_i)
\right\}
\right].
\end{aligned}
\end{equation}
\end{definition}

The probability in \eqref{eq:payoff-perturbed-ohe-choice} averages over the
subject's entire private shock vector, holding $\beta_i$ fixed. Partners'
shocks enter through their aggregate action distributions and consistency;
independence makes a subject's own shocks uninformative about the partner's
action. The definition averages optimal choices made \emph{after} subjects
observe their private shocks.

Call a profile stationary at a fixed $\varepsilon$ if, initialized with
that profile as the preceding action distribution, the learning process
reproduces the same population shares in every period.

\begin{proposition}
\label{prop:learning-foundation}
For every $\varepsilon>0$, the stationary aggregate profiles of the
specified learning process coincide with the payoff-perturbed OHE profiles
at scale $\varepsilon$, and this set is nonempty. Every accumulation point
of such profiles as $\varepsilon$ tends to zero is an OHE of the
unperturbed environment.
\end{proposition}

The proof is in Appendix~\ref{proof:learning-foundation}. At a stationary profile, the
conjectures based on the preceding period's action frequencies are consistent
with current behavior, while population action shares are induced by optimal
choices under private perturbations. As the perturbations vanish, any
advantage from deviating in the unperturbed decision problem vanishes as
well. The result concerns stationary profiles and their limits across
perturbation scales. It does not establish convergence of learning over
time, or that every OHE is obtained under the chosen shock specification;
Online Appendix~\ref{oa:learning-scope} gives a counterexample to the converse.

The same grouping of roles gives an exact
analogy-based expectation equilibrium representation of unperturbed OHE
under fixed matching: one class for each other ingroup role and one
class for the whole outgroup, when nonempty. Section~\ref{sec:analogy-partitions}
uses this representation to discuss alternative partitions.

\subsection{Random-matching benchmark}\label{sec:benchmark-matching}
Fix the characteristic profile \(x\) and the groups. Consider the family
of games \((\Gamma^{x,\varphi})_{\varphi\in\Phi}\) described in
Section~\ref{section:strategic_environment}. A matching is drawn according
to a distribution \(\mu\in\Delta(\Phi)\). If the draw is \(\varphi\),
player \(i\) interacts with \(\varphi(i)\) and the outcome function is
\(f^{x,\varphi}\). Before choosing an action, player \(i\) receives a
signal: an ingroup match reveals the partner's identity, whereas an outgroup
match is reported only as such. The family of games, the characteristic
profile, the groups, the distribution \(\mu\) and the signal functions are
common knowledge. Thus \(\mu\) is the common prior over matchings in this Bayesian game. On its
support the signal of player \(i\) is
\[
\tau_i(\varphi)=
\begin{cases}
\varphi(i),&\text{if }\varphi(i)\in G(i),\\
\mathrm{Out},&\text{if }\varphi(i)\notin G(i),
\end{cases}
\]
where \(\mathrm{Out}\) is distinct from player identities. Players receive no further signal before acting.

Let \(T_i=\{\tau_i(\varphi):\varphi\in\operatorname{supp}\mu\}\) be the set of possible signals. A mixed decision function \(s_i:T_i\to\Delta(A)\) specifies the mixed action \(s_i(t_i)\) chosen after each signal \(t_i\). Players act simultaneously and, conditional on \(\varphi\), their actions are drawn independently with distributions \(s_i(\tau_i(\varphi))\).

Given a profile \(s=(s_i)_{i\in I}\), write \(s_{-i}\) for the other players' decision functions. Player \(i\)'s conditional expected payoff from choosing \(\sigma_i'\in\Delta(A)\) after \(t_i\in T_i\) is
\[
\bar U_i(\sigma_i',s_{-i};t_i)
=
\sum_{\substack{\varphi\in\operatorname{supp}\mu\\
                \tau_i(\varphi)=t_i}}
\mu(\varphi\mid t_i)\,
U_i\!\left(
\sigma_i',
s_{\varphi(i)}\bigl(\tau_{\varphi(i)}(\varphi)\bigr);
x_i,x_{\varphi(i)}
\right),
\]
where \(\mu(\varphi\mid t_i)\) is the prior conditioned on player \(i\)'s signal. The expectation retains each possible partner's characteristics and her action distribution at the signal she receives.

A profile \(s\) is a \emph{Bayes--Nash equilibrium} (BNE), under the prior \(\mu\), if for every \(i\in I\), \(t_i\in T_i\), and \(\sigma_i'\in\Delta(A)\),
\[
\bar U_i(s_i(t_i),s_{-i};t_i)
\ge
\bar U_i(\sigma_i',s_{-i};t_i).
\]
In equilibrium, each player chooses optimally conditional on her signal and correctly anticipates how each possible partner plays given the signal that partner receives. In an ingroup match with \(j\), player \(i\)'s conjecture about her partner's action equals \(s_j(i)\), since \(j\) receives signal \(i\). After \(\mathrm{Out}\), each possible partner also receives \(\mathrm{Out}\). Player \(i\)'s conjecture is then the average of the mixed actions \(s_j(\mathrm{Out})\), weighted by the posterior probability of meeting each \(j\). This is the key difference with the OHE notion: for an outgroup match, the OHE consistency requirement pools actual behavior over the entire outgroup, including members who are themselves matched ingroup.

\paragraph{Equivalence of equilibrium choices.}
When do equilibrium choices in the random-matching benchmark coincide
with those under OHE at a fixed matching? The following result gives
sufficient conditions.

Suppose \(|G_1|=|G_2|=n\ge1\), so exclusively outgroup matchings are feasible, and let
\[
\Phi_{\mathrm{Out}}
=
\{\varphi\in\Phi:\varphi(i)\notin G(i)
\text{ for every }i\in I\}.
\]
Under the uniform prior \(\mu\) on this set, \(T_i=\{\mathrm{Out}\}\) for every \(i\). We therefore identify each decision function with the mixed action \(\sigma_i=s_i(\mathrm{Out})\).

The strategic environment has \emph{private values} if, for every \(i\in I\), \(
u_i(a_i,a_j;x_i,x_j)
=
u_i(a_i,a_j;x_i,x_j')
\) for all \(a_i,a_j\in A\) and all \(x_i,x_j,x_j'\in X\).
Thus, holding both actions fixed, a player's payoff does not
depend on her partner's characteristics (although those characteristics
may still affect the partner's action).

\begin{proposition}
\label{prop:ohe-bne-equivalence}
Suppose the strategic environment has private values and
\(|G_1|=|G_2|=n\ge1\). Let \(\mu\) be the uniform distribution on
\(\Phi_{\mathrm{Out}}\). For every \(x\in X^I\), every
\(\varphi\in\Phi_{\mathrm{Out}}\), and every profile
\(\sigma\in(\Delta(A))^I\), \(\sigma\) is an OHE under the fixed
matching \(\varphi\) if and only if the decision functions
\(s_i(\mathrm{Out})=\sigma_i\), \(i\in I\), form a BNE under
\(\mu\).
\end{proposition}

Under these conditions, a player who observes her outgroup partner and
uses the group aggregate faces the same optimization problem as a player in the random-matching benchmark (Appendix~\ref{proof:random-equivalence}). In the benchmark her signal is always
$\mathrm{Out}$, and each member of the opposite group is equally likely
to be her partner. In particular, the OHE profile
set does not depend on which matching in \(\Phi_{\mathrm{Out}}\)
is fixed.

This set of conditions is sufficient for equivalence, but need not be
necessary in every game. By contraposition, different equilibrium
action-profile sets require at least one condition to fail.
The coordination illustration departs from private values,
whereas the organizational application allows both ingroup and outgroup matches.
Online Appendix~\ref{oa:bounded-effort-bayesian} establishes the corresponding equivalence for bounded effort under full outgroup matching.

%%%%%%%%%%%%%%%%%%
%%%    ORGS    %%%
%%%%%%%%%%%%%%%%%%
\section{An Application to Organizations}\label{section:orgs}

Organizations that combine individual performance evaluation within units
with aggregate reporting across units provide a natural setting for the
application. Distinct units can carry out comparable activities, as in
firms that replicate a common operating model across outlets
\citep{WinterSzulanski2001}. Individual appraisals evaluate workers
separately \citep{DeNisiPritchard2006}, whereas unit-level key performance
indicators (KPIs) summarize collective results and communicate them beyond
the unit \citep{KaplanNorton1996}. Direct contact across unit boundaries
can also be limited \citep{Hansen1999}, while departments develop shared
ways of interpreting organizational activity that differ across units
\citep{Dougherty1992}. Together, these features motivate a setting in
which workers form differentiated assessments of colleagues within their
own unit and shared assessments of workers in other units.%
\footnote{This interpretation connects to accounts of organizational culture
as shared meaning and interpretation
\citep{Pettigrew1979,GibbonsPrusak2020}, which can differ across departments
\citep{Dougherty1992}. \citet{GibbonsLiCalziWarglien2021} formalize how
shared cognitive frames shape the categorization of situations and
collective performance. These perspectives motivate viewing shared
outgroup assessments as an aspect of unit culture; the application's
conjecture restrictions specify their form.}

This application considers a simple setting of moral hazard in
teams \citep{Holmstrom1982}, where teammates' efforts are strategic
complements. A designer assigns workers to identical two-person tasks,
taking their unit membership as given. Workers form partner-specific
expectations within their own unit and assess outgroup teammates through
the behavior of the partner's unit as a whole, including colleagues
working in internal teams. The workers who collaborate across units need
not be representative of this wider group. Team formation therefore
determines both who works together and the expectations guiding their
behavior. The designer chooses team assignments to maximize expected task
success, keeping the bonus schedule fixed.

\paragraph{Environment.}

The population \(I\) is nonempty and performs \(|I|/2\) identical
two-agent tasks, exhibiting strategic complementarity. The groups represent fixed organizational units.
Agent \(i\) has an effort-cost coefficient \(x_i>0\) and chooses effort
\(e_i\in A=[0,\bar e]\), where \(\bar e>0\), at cost \(x_i e_i^2/2\).

The probability that a task undertaken by agents exerting efforts \(e\)
and \(e'\) succeeds is
\begin{equation}
\label{eq:org-success}
p(e,e')=\kappa(e+e'+t ee').
\end{equation} 
where \(0<\kappa\le (2\bar e+\bar e^2)^{-1}\) is a constant ensuring \(0\le p(e,e')\leq 1\) and the parameter \(t\in(0,1)\) captures the intensity of strategic
complementarity: on \(A^2\), 
\(\partial^2p/\partial e\partial e'=\kappa t>0\).

Effort is non-contractible, whereas task success is contractible. Both teammates receive a unit bonus if their task succeeds and zero otherwise. Agents are risk neutral, so the expected payoff of agent \(i\) matched with \(j\) is
\begin{equation}
\label{eq:org-payoff}
u_i(e_i,e_j;x_i,x_j)
=\kappa(e_i+e_j+t e_i e_j)-\frac{x_i e_i^2}{2}.
\end{equation}
The partner's cost affects this payoff only through the partner's effort, so this is a setting with private values.

The designer knows the cost profile and technology and chooses a matching,
holding groups and the bonus schedule fixed. At the action stage, each
worker knows her payoff rule and observes her teammate's identity, both
cost coefficients, and both group memberships, as in the general framework. Agents choose efforts simultaneously; task outcomes then
determine their bonus payments.

\paragraph{Effort choices and equilibrium.}

Since effort is chosen in an interval of the real line, \(\Delta(A)\) denotes the Borel probability measures on \(A\).
The definition of OHE applies with the mixed-payoff extension
\[
U_i(\alpha,\gamma;x_i,x_j)
=\int_A\int_A u_i(e,e';x_i,x_j)\,\gamma(de')\,\alpha(de),
\qquad \alpha,\gamma\in\Delta(A).
\]
Continuity and boundedness of the pair payoff ensure that this integral
is finite. For a conjecture \(\beta_i\), let
\(m_i=\int_A s\,\beta_i(ds)\). Then
\[
U_i(\delta_e,\beta_i;x_i,x_{\varphi(i)})
=\kappa\bigl[e+(1+t e)m_i\bigr]-\frac{x_i e^2}{2}.
\]
Connectedness of the effort set ensures that \(m_i\) is a feasible effort,
while affinity of the payoff in the partner's effort makes \(\beta_i\)
payoff-equivalent to the deterministic conjecture \(\delta_{m_i}\), for
every own pure or mixed action. This payoff equivalence leaves the
distributional consistency requirement unchanged.

Write \(c_i=x_i/\kappa\) for the normalised cost coefficient. Strict
concavity in own effort gives the unique best response
\[
\min\left\{\bar e,\frac{1+t m_i}{c_i}\right\}.
\]
Every mixed best response assigns probability one to this effort.
For a pure effort profile \(e\), consistency determines the conjectured
mean
\begin{equation}
\label{eq:org-means-map}
m_i(e)=
\begin{cases}
e_{\varphi(i)},
&\varphi(i)\in G(i),\\[3pt]
\displaystyle\frac{1}{|G(\varphi(i))|}
\sum_{j\in G(\varphi(i))}e_j,
&\varphi(i)\notin G(i).
\end{cases}
\end{equation}
The target group in the second line contains the actual partner and is
therefore nonempty. An effort profile represents an OHE if and only if
\begin{equation}
\label{eq:org-capped}
e_i=\min\left\{\bar e,\frac{1+t m_i(e)}{c_i}\right\}
\qquad\text{for every }i\in I.
\end{equation}

\begin{proposition}
\label{prop:org-ohe}
For every positive cost profile and every matching \(\varphi\in\Phi\),
there exists a unique OHE, which is pure.
\end{proposition}

Existence follows from Brouwer's theorem applied to the continuous capped
best-response map. For uniqueness, comparing the largest coordinate-wise
effort ratio between two candidate equilibria yields a contradiction:
conjectured means are linear in efforts, while uncapped responses have a
strictly positive constant term. Strict concavity ensures that equilibrium
behavior is pure. The proof is in Appendix~\ref{proof:org-ohe}.

Denote the equilibrium effort profile by \(e(\varphi)\). For the following
characterization, suppose that
\begin{equation}
\label{eq:org-interiority}
c_i>t+\frac1{\bar e}
\qquad\text{for every }i\in I,
\end{equation}
so that all equilibrium efforts are strictly interior. For each nonempty
group \(G\), let
\[
\eta_G(\varphi)=\frac1{|G|}\sum_{i\in G}e_i(\varphi)
\]
denote its average equilibrium effort. Then, for every \(i\in G\),
\begin{equation}
\label{eq:org-efforts}
e_i(\varphi)=
\begin{cases}
\displaystyle\frac{c_{\varphi(i)}+t}
{c_i c_{\varphi(i)}-t^2},
&i\in\operatorname{In}_G^\varphi,\\[8pt]
\displaystyle\frac{1+t\eta_{G(\varphi(i))}(\varphi)}{c_i},
&i\in\operatorname{Out}_G^\varphi.
\end{cases}
\end{equation}
At fixed \(t\), the ingroup expression is strictly decreasing and jointly
strictly convex in the two partners' costs. The outgroup expression is
strictly decreasing and strictly convex in own cost holding the group
mean fixed, and increasing and affine in that mean.

When both groups are nonempty, their averages uniquely solve the following
two linear equations, where \(G'\) denotes the other group:
\begin{equation}
\label{eq:org-means}
|G|\eta_G(\varphi)
=\sum_{i\in\operatorname{In}_G^\varphi}e_i(\varphi)
+\bigl(1+t\eta_{G'}(\varphi)\bigr)
\sum_{i\in\operatorname{Out}_G^\varphi}\frac1{c_i},
\quad G\in\{G_1,G_2\}.
\end{equation}
The ingroup efforts are given by the first line of
\eqref{eq:org-efforts}. If one group is empty, every agent's effort is
given by the ingroup formula.

Throughout this interior domain, for a fixed matching and fixed \(t\),
each group average is nonincreasing and jointly convex in the cost
profile. Cost monotonicity is strict for own-group costs and, whenever
outgroup tasks exist, for every agent's cost. Holding costs fixed, each
average is strictly increasing and strictly convex in \(t\).

To see these properties, fix \(\varphi\) and let \(w_{ij}\ge0\) denote the
weight on agent \(j\)'s effort in agent \(i\)'s consistent mean. These
weights are determined by the matching and groups and satisfy
\(\sum_jw_{ij}=1\). Repeated substitution of the interior effort equations
gives
\begin{equation}
\label{eq:org-series}
\begin{aligned}
e_i(\varphi)
&=\frac1{c_i}+\frac{t}{c_i}\sum_jw_{ij}e_j(\varphi)\\
&=\frac1{c_i}
+t\sum_j\frac{w_{ij}}{c_ic_j}
+t^2\sum_{j,k}\frac{w_{ij}w_{jk}}{c_ic_jc_k}
+\cdots .
\end{aligned}
\end{equation}
The series converges because \(\max_i t/c_i<1\). At fixed \(t\), each
term is nonincreasing and jointly convex in costs: the cost dependence
consists of products of inverse costs with nonnegative coefficients.
Holding costs fixed, the strictly positive coefficients of \(t\) and
\(t^2\) imply that effort is strictly increasing and strictly convex in
\(t\). Averaging this expansion over the members of a group establishes
the corresponding properties of \(\eta_G(\varphi)\). The detailed proof
and explicit solution for the averages appear in Appendix~\ref{proof:org-interiority}.

\par\addvspace{\topsep}\noindent\begin{minipage}{\linewidth}
\begin{example}[Effort across three tasks]\label{ex:org-effort-spillover}
\normalfont
Each group has two workers of cost \(c_L\) and one of cost \(c_H>c_L\),
with both costs satisfying~\eqref{eq:org-interiority}. Pair the two
low-cost workers in each group together and the two high-cost workers
across groups. Each low-cost worker exerts \(e_L=1/(c_L-t)\), whether
under OHE or Nash equilibrium. What changes is the effort of the
high-cost pair. Under OHE, each expects her teammate to exert the other
group's mean effort,
\(\eta=(2e_L+e_H)/3\). Substituting this mean into
\(e_H=(1+t\eta)/c_H\) gives
\[
  e_H=\frac{3+2t e_L}{3c_H-t},
  \qquad
  e_H-\frac{1}{c_H-t}
  =\frac{2t}{3c_H-t}\left(e_L-\frac{1}{c_H-t}\right)>0 .
\]
Under Nash equilibrium each high-cost worker responds to her actual
teammate and exerts \(1/(c_H-t)\). Under OHE, the effort of the two
low-cost workers in her teammate's group raises the effort she expects
from her teammate, even though she cannot meet either low-cost worker at this matching.
\end{example}
\end{minipage}\par\addvspace{\topsep}

\paragraph{The designer's problem.}
With groups and the bonus schedule fixed, the designer chooses a matching
\(\varphi\in\Phi\) to maximise the expected number of successful tasks,
\begin{equation}
\label{eq:org-objective}
S(\varphi)
=
\frac{1}{2}\sum_{i\in I}
p\bigl(e_i(\varphi),e_{\varphi(i)}(\varphi)\bigr).
\end{equation}
The factor \(1/2\) ensures that each task is counted once. Success is
evaluated using the efforts of the actual matched partners.

For any fixed matching, the expected number of successful tasks is
supermodular in efforts: the cross derivative equals \(\kappa t\) for
teammates and zero for workers on different tasks.\footnote{A function
\(f:D\to\mathbb R\), where \(D\) is a product of nondegenerate intervals,
is supermodular if
\[
f(\max\{z,z'\})+f(\min\{z,z'\})
\geq f(z)+f(z')
\qquad\text{for all }z,z'\in D,
\]
where maximum and minimum are taken coordinatewise. For a twice
continuously differentiable function, this is equivalent to nonnegative
cross partial derivatives between distinct coordinates. See
\citet{BattigalliCatoniniDeVito2025}.}
Under the interiority condition~\eqref{eq:org-interiority}, the objective
evaluated at equilibrium is also supermodular in costs.\footnote{
Supermodularity in costs need not hold when effort caps bind, while
supermodularity in efforts holds throughout the feasible effort domain.}

To see this, recall that the equilibrium effort expansion above expresses
each equilibrium effort as a convergent sum of products of inverse powers
of costs, with nonnegative coefficients. Since task success combines
efforts and their products with positive coefficients, substituting these
expansions into~\eqref{eq:org-objective} gives a series of the same form.
Each inverse-cost product is supermodular, and nonnegative sums and
pointwise limits preserve this property. Appendix~\ref{proof:org-supermodularity} gives the supermodularity argument; Online Appendix~\ref{oa:strict-cost-effects} characterizes when the cross-cost effects are strict.

These complementarities also guide comparisons across matchings. Holding
efforts fixed, supermodularity favors pairing high-effort workers
together. For ingroup teams, the same expansion establishes complementarity
in teammates' costs for both task success and total effort. These
properties help evaluate the direct gains from reassignment and, through
changes in group averages, their effects on outgroup effort. Consider
first the choice of partners, holding fixed which workers undertake
ingroup and outgroup tasks.

\begin{definition}
\label{def:org-one-sided-assortative}
\label{def:org-two-sided-assortative}
A matching \(\varphi\) is \emph{one-sided assortative in \(J\subseteq I\)},
where \(J\) contains \(2m\) workers, if its members can be listed in
nondecreasing order of cost as \(i_1,\ldots,i_{2m}\), with
\(\varphi(i_{2r-1})=i_{2r}\) for \(r=1,\ldots,m\).

It is \emph{two-sided assortative between disjoint subsets
\(J_1,J_2\subseteq I\)}, each containing \(m\) workers, if their members
can be listed separately in nondecreasing order of cost as
\(i_1,\ldots,i_m\) and \(j_1,\ldots,j_m\), with
\(\varphi(i_r)=j_r\) for \(r=1,\ldots,m\).
\end{definition}

When costs tie, it suffices that some ordering of the tied workers
satisfies the relevant requirement. Both definitions apply vacuously when
the corresponding subsets are empty.

Fix the ingroup and outgroup subsets \(\operatorname{In}_G\) and
\(\operatorname{Out}_G\) induced by some matching, allowing empty subsets.

\begin{proposition}
\label{prop:org-conditional-sorting}
Suppose the interiority condition~\eqref{eq:org-interiority} holds.
Among matchings inducing these subsets, a matching maximises the expected
number of successful tasks if and only if it is one-sided assortative in
each \(\operatorname{In}_G\) and two-sided assortative between
\(\operatorname{Out}_{G_1}\) and \(\operatorname{Out}_{G_2}\).
\end{proposition}

For ingroup tasks, complementarity in costs makes assortative pairing increase both task success and total effort. Higher total effort also encourages outgroup effort through group averages. For outgroup tasks, changing partners leaves equilibrium efforts unchanged, so complementarity in efforts favors pairing workers with corresponding effort ranks. Since effort decreases with cost within each outgroup subset, this also means pairing corresponding cost ranks.\footnote{Without interiority, cost-assortative outgroup pairing remains optimal for fixed ingroup and outgroup subsets and fixed ingroup pairings. Outgroup reassignment still preserves equilibrium efforts, which remain weakly decreasing in cost. Caps can give workers with different costs identical efforts, making other cost pairings equally good. Ingroup cost assortativity, however, need not remain optimal when caps bind.}
The proposition thus determines how to pair workers conditional on the ingroup and outgroup subsets (Appendix~\ref{proof:org-conditional-sorting} gives the proof). It leaves open which workers should be matched ingroup and which outgroup.

\subsection{Changing who works outgroup: local rematches}

The smallest change in a matching involves two tasks and four workers. Starting with one ingroup task in each group, the designer can form two outgroup tasks by pairing each worker with a teammate from the other group. Conversely, starting with two outgroup tasks, the designer can pair the two workers from each group together. All other workers keep their teammates, but those working outgroup may still adjust their effort if the rematch changes average effort in their teammate's group. A matching can be optimal only if no such rematch raises the expected number of successful tasks.

\paragraph{From ingroup to outgroup tasks.}
Consider first a matching \(\psi\) that replaces two ingroup tasks under \(\phi\) with two outgroup tasks. Let \(i_k,j_k\) be the original teammates in \(G_k\), with \(c_{i_k}\leq c_{j_k}\). Under \(\psi\), \(i_1\) and \(i_2\) work together, as do \(j_1\) and \(j_2\). Suppose the two cost intervals overlap and both original group averages are at least as high as the largest of the four original efforts:
\[
\max\{c_{i_1},c_{i_2}\}\leq\min\{c_{j_1},c_{j_2}\},
\qquad
\max\{e_{i_1}(\phi),e_{i_2}(\phi)\}
\leq\min\{\eta_{G_1}(\phi),\eta_{G_2}(\phi)\}.
\]
Here coarse conjectures and supermodularity favor the same rematch. Even if all four workers best-replied to the smaller of the two original group averages, each would exert at least as much effort as with the former teammate. At that common average, effort decreases with own cost. Overlap therefore ensures that the new pairing puts the two highest efforts on the same task, as supermodularity favors. In equilibrium, all four workers weakly increase effort, both group averages weakly rise, and the rematch strictly raises the expected number of successful tasks unless all four original efforts and both original group averages coincide. The argument is in Online Appendix~\ref{app:local-rematches}.

How much of this worker-by-worker improvement can the designer dispense with while both group averages still rise? First hold the averages at their values under \(\phi\). A worker \(i\in G_k\) moving outgroup would then exert \((1+t\eta_{G_\ell}(\phi))/c_i\), where \(G_\ell\) is the other group. One former teammate may exert less while the two together exert weakly more. If this holds in each group, that is, if
\[
(1+t\eta_{G_\ell}(\phi))
\left(\frac{1}{c_{i_k}}+\frac{1}{c_{j_k}}\right)
\geq e_{i_k}(\phi)+e_{j_k}(\phi),
\qquad k=1,2,\quad \ell\ne k,
\]
then both equilibrium averages weakly rise under \(\psi\) by~\eqref{eq:org-means}. This encourages effort on every outgroup task. The designer must still account for how the new teams combine those efforts. Either way of pairing the four workers across groups induces the same equilibrium efforts, yet the two pairings can yield different expected numbers of successful tasks.

To compare the new tasks with the original ones, let \(g_k=\sqrt{e_{i_k}(\phi)e_{j_k}(\phi)}\) denote the geometric mean of the original teammates' efforts in \(G_k\). Its square is the effort product that enters their task's success probability. When the cost intervals overlap, the change in expected successes on the two selected tasks, evaluated at the original group averages, has the lower bound \(\underline{\Delta S}_{\mathrm{direct}}(\phi)\) given by
\[
\begin{aligned}
\frac{\underline{\Delta S}_{\mathrm{direct}}(\phi)}{\kappa t}
={}&
(1+t\eta_{G_1}(\phi))(1+t\eta_{G_2}(\phi))
\left(\frac{1}{c_{i_1}}-\frac{1}{c_{j_2}}\right)
\left(\frac{1}{c_{i_2}}-\frac{1}{c_{j_1}}\right)\\
&+2\sum_{k=1}^{2}g_k
\left(\sqrt{\eta_{G_1}(\phi)\eta_{G_2}(\phi)}-g_k\right).
\end{aligned}
\]
The first line captures the contribution from pairing the lower-cost members of the two original teams together. Overlap makes this term nonnegative. The second line shows what this gain may have to cover. A team whose original effort product was larger carries greater weight in the comparison. If \(\underline{\Delta S}_{\mathrm{direct}}(\phi)\geq0\), the two new tasks together yield at least as many expected successes as the original tasks even at the original group averages. If both averages then weakly rise, the adjustment to the new equilibrium preserves this gain.

How far can one group's average effort fall while the rematch still raises the expected number of successful tasks? The direct comparison already accounts for breaking up the original teams and forming the two outgroup tasks at the original averages. It remains to account for the change in conjectures on every outgroup task under \(\psi\), including tasks whose workers keep their teammates. Workers who remain paired ingroup continue to best-reply to each other and exert the same efforts under both matchings.

\begin{proposition}
\label{prop:rematch-falling-average}
Suppose condition~\eqref{eq:org-interiority} holds and the cost intervals of the two selected ingroup pairs overlap. Let \(\psi\) pair \(i_1\) with \(i_2\) and \(j_1\) with \(j_2\), keeping all other teammates fixed. If \(\eta_{G_1}(\psi)<\eta_{G_1}(\phi)\) and \(\eta_{G_2}(\psi)\geq\eta_{G_2}(\phi)\), then \(S(\psi)>S(\phi)\) whenever
\[
\eta_{G_1}(\phi)-\eta_{G_1}(\psi)
\;\leq\;
\frac{
\displaystyle
\frac{\underline{\Delta S}_{\mathrm{direct}}(\phi)}{\kappa t}
+
\bigl[\eta_{G_2}(\psi)-\eta_{G_2}(\phi)\bigr]
\sum_{i\in\mathrm{Out}_{G_1}^{\psi}}
\frac{1}{c_i}
\left(1+t\,\frac{1+t\eta_{G_1}(\phi)}{c_{\psi(i)}}\right)
}{
\displaystyle
\sum_{j\in\mathrm{Out}_{G_2}^{\psi}}
\frac{1+t e_{\psi(j)}(\psi)}{c_j}
}.
\]
The same conclusion holds with the groups interchanged.
\end{proposition}

To read the bound, first hold \(G_1\)'s average at its original value and raise \(G_2\)'s average to its value under \(\psi\). Workers in \(\mathrm{Out}_{G_1}^{\psi}\) exert more as their conjectured mean rises. By supermodularity, each increase contributes more to task success when the worker's actual teammate exerts greater effort. This gives the second term in the numerator. Now let \(G_1\)'s average fall. Workers in \(\mathrm{Out}_{G_2}^{\psi}\) reduce their efforts, and the denominator measures the resulting loss per unit of decline, in units of \(\kappa t\). The gain on the two rematched tasks can be large enough that the designer prefers \(\psi\) even when total effort across the organization falls. Appendix~\ref{proof:opening-bound} proves the bound; Online Appendix~\ref{oa:opening-example} gives an example in which total effort falls.

\paragraph{From outgroup to ingroup tasks.}
Consider two outgroup tasks under \(\phi\). Let \(\psi\) pair the two selected workers in each group together, leaving all other teammates fixed. Suppose both selected workers in \(G_1\) have strictly lower costs than either selected worker in \(G_2\). Under \(\psi\), the two lower-cost workers best-reply to each other, and supermodularity makes their joint effort especially productive. If \(\eta_{G_k}(\psi)\geq\eta_{G_k}(\phi)\) for \(k=1,2\), the rematch strictly raises the expected number of successful tasks (Online Appendix~\ref{proof:closing-separated}). This holds even when both averages stay fixed. Total effort then stays fixed as well, and the gain comes entirely from the effort products on the two rematched tasks. An optimal \(\phi\) can therefore keep these four workers outgroup only if bringing them ingroup would lower average effort in at least one group. Even then, a sufficiently small decline can leave the rematch profitable, as the example in Online Appendix~\ref{oa:closing-example} shows.

\paragraph{Fixed number of outgroup matches.}
A local rematch can also change which worker works outgroup while keeping the number of outgroup tasks fixed. Suppose that under \(\phi\), workers \(i\) and \(j\) are ingroup teammates and a third worker \(k\) in their group works outgroup. Let \(\psi_i\) exchange the places of \(i\) and \(k\): worker \(i\) takes \(k\)'s outgroup teammate \(\phi(k)\), and \(k\) works with \(j\). Define \(\psi_j\) analogously. Both rematches keep all other teammates fixed.

\begin{proposition}[Consecutive ranks of ingroup teammates]
\label{prop:consecutive-ingroup-ranks}
Fix \(\phi\) and workers \(i,j,k\) as above. Under~\eqref{eq:org-interiority},
\[
c_i<c_k<c_j
\quad\Longrightarrow\quad
S(\phi)<\max\{S(\psi_i),S(\psi_j)\}.
\]
Consequently, at any matching that admits no strictly profitable two-task rematch, and hence at every optimal matching, each group has a nondecreasing ordering of all its workers by cost in which every ingroup pair occupies consecutive ranks.
\end{proposition}

Proposition~\ref{prop:org-conditional-sorting} places ingroup teammates consecutively when only the ingroup workers are ordered by cost. Proposition~\ref{prop:consecutive-ingroup-ranks} extends this property to the whole group, allowing a suitable ordering of workers whose costs coincide. An outgroup worker can still lie between two different ingroup teams. For instance, two workers with costs below hers could work together, and two workers with costs above hers could also work together. Thus, even when the number of outgroup workers is fixed, the proposition leaves open whether they are the highest-cost workers in the group. The proof is in Appendix~\ref{proof:ingroup-adjacency}.

With three distinct costs and one outgroup worker in a group, the proposition
rules out sending the middle-ranked worker: doing so would leave the two
extremes paired ingroup. In Example~\ref{ex:org-effort-spillover} the
two low costs coincide, so adjacency alone does not decide whom to send.
There the illustrated matching is in fact the unique optimum. It keeps
each low-cost pair together and lets the two high-cost workers respond to
the resulting group means. Online Appendix~\ref{proof:six-optimum} compares the feasible matchings.

Consecutive ranks nevertheless have a direct implication for large groups. Suppose a group has \(n\) workers with costs in a fixed interval \([a,b]\). At an optimal matching, the cost intervals spanned by its ingroup pairs have disjoint interiors, so their lengths sum to at most \(b-a\). For any \(\varepsilon>0\), at most \((b-a)/\varepsilon\) ingroup pairs can therefore have a cost gap of at least \(\varepsilon\). The fraction of the group whose ingroup teammate differs in cost by at least \(\varepsilon\) is at most \(2(b-a)/(n\varepsilon)\), which vanishes as the group grows. This makes the effort a worker of cost \(c\) would exert with an equally costly ingroup teammate, \(1/(c-t)\), the natural benchmark for the next comparison.

\subsection{A common cost cutoff in large finite organizations}
\label{subsec:large-finite-organizations}

Proposition~\ref{prop:consecutive-ingroup-ranks} shows that, as a group grows, the fraction of its workers whose ingroup teammate differs appreciably in cost vanishes. Which workers, then, should work ingroup? When the two groups approach the same cost distribution, a common cutoff answers this question for almost all workers in every optimal matching.

Consider a sequence of organizations indexed by \(n\). Organization \(n\) consists of two disjoint groups \(G_{1,n}\) and \(G_{2,n}\), each with \(n\) workers, who together carry out \(n\) two-person tasks. The parameters \(t\), \(\kappa\), and \(\bar e\) remain fixed. In every organization, the normalized costs \(c_i\) lie in the same interval \([a,b]\), with \(t+1/\bar e<a<b\), so condition~\eqref{eq:org-interiority} holds throughout. Suppose the empirical cost distribution in each group converges weakly to the same absolutely continuous distribution \(F\) on \([a,b]\). Since \(F\) is continuous, this means that, for each \(k=1,2\) and every \(c\in[a,b]\),
\[
\frac{1}{n}\bigl|\{i\in G_{k,n}:c_i\leq c\}\bigr|
\;\longrightarrow\; F(c).
\]
The two groups may have different cost lists at every finite \(n\). The symmetry assumption concerns their limiting distributions.

A worker of cost \(c\) exerts \(1/(c-t)\) with an equally costly ingroup teammate. If she works outgroup and the average effort in her teammate's group is \(\eta\), she exerts \((1+t\eta)/c\). She therefore exerts more ingroup precisely when her equally costly teammate would exert more than the opposite group's average. For a given \(\eta>0\), the two efforts coincide at \(c=t+1/\eta\). To determine the cutoff, the average must itself agree with the efforts induced by the assignment. Let \(\eta^*\) be the unique solution to
\[
\eta^*
=\int_a^b
\max\left\{\frac{1}{c-t},\,\frac{1+t\eta^*}{c}\right\}\,dF(c),
\]
and write \(c^*=t+1/\eta^*\). The integral averages over all workers, including those who work ingroup, as required by the outgroup conjecture.

\begin{proposition}[Common cost cutoff]
\label{prop:large-finite-cutoff}
Let \(\phi_n\) be an optimal matching in organization \(n\). As \(n\) grows, the same cutoff \(c^*\) describes the assignment of all but a vanishing fraction of workers in each group: workers with costs below \(c^*\) work ingroup, and workers with costs above \(c^*\) work outgroup. The cutoff lies strictly between the lowest and highest costs in the support of \(F\). In each group \(k=1,2\),
\[
\eta_{G_{k,n}}(\phi_n)\longrightarrow\eta^*,
\qquad
\frac{\bigl|\mathrm{Out}_{G_{k,n}}^{\phi_n}\bigr|}{n}
\longrightarrow 1-F(c^*)\in(0,1).
\]
The latter limit is also the fraction of tasks carried out by outgroup teammates in the limit.
\end{proposition}

To see why the effort comparison identifies an optimal assignment, consider a feasible matching that pairs consecutive low-cost workers within each group and pairs the remaining higher-cost workers across groups in cost order. The designer can choose a common number of outgroup workers close to \(n[1-F(c^*)]\), leaving an even number in each group to pair ingroup. Consecutive ingroup teammates then have similar costs on almost all tasks. Because the groups approach the same cost distribution, the same holds for outgroup teammates at corresponding ranks. This matching attains, in the limit, an upper bound on expected successes per task. Assigning a positive fraction of workers to the other type of task at costs bounded away from \(c^*\) would leave a strictly positive loss per task. This is why the cutoff describes every sequence of optimal matchings.

Appendix~\ref{proof:large-finite-cutoff} proves the cutoff result and gives the limits of effort and task success. In particular, writing
\[
e^*(c)=\max\left\{\frac{1}{c-t},\,\frac{1+t\eta^*}{c}\right\},
\]
the average absolute difference between \(e_i(\phi_n)\) and \(e^*(c_i)\) tends to zero in each group, and expected successes per task converge to
\(\int_a^b p(e^*(c),e^*(c))\,dF(c)\).
The same conclusions, including the cutoff prediction, hold for any sequence of matchings whose shortfall from optimal expected successes per task tends to zero.

\paragraph{Computing the cutoff and moving it.}
Given the limiting cost distribution, the designer can compute \(c^*\)
by iterating the mean equation. Start with the mean effort from
equal-cost ingroup pairing, \(\int_a^b (c-t)^{-1}\,dF(c)\), and
replace each trial mean by the right-hand side of the equation. The
first step weakly raises the mean because the maximum in the integral
is at least the ingroup effort at each cost. The right-hand side is
increasing in the trial mean and has Lipschitz constant at most
\(t/a<1\). Thus the sequence of means rises weakly to \(\eta^*\),
and the implied cutoffs fall weakly to \(c^*\). This computes the
limiting cutoff from the cost distribution without having to try
successive matchings in a finite organization.

At fixed technology, a higher equilibrium mean lowers \(c^*\):
outgroup work elicits more effort at a wider range of costs. For
example, consider a mean-preserving spread of the common limiting cost
distribution within the maintained domain. At each trial mean, both
effort branches are convex in cost, and so is their maximum. The
spread therefore weakly increases the right-hand side of the mean
equation at the original fixed point. Iteration from that point gives
a weakly higher new fixed point and hence a weakly lower cutoff.
To determine the change in the outgroup share, one must also account
for how the cost distribution has moved relative to the cutoff.

\paragraph{Comparison with random matching and Nash equilibrium.}
When every task is ingroup, OHE agrees with Nash at the given matching.
When every task is outgroup, individual efforts under OHE coincide with those in the random-matching benchmark
when all full-outgroup matchings are equally likely and each worker
receives only the signal \(\mathrm{Out}\) (Online Appendix~\ref{oa:bounded-effort-bayesian}). This agreement concerns choices:
actual task success can still differ because success depends on which
efforts are paired together.

Partial segregation can break the agreement even in
Example~\ref{ex:org-effort-spillover}. Suppose workers receive no signal
about the matching. Under any common prior on perfect matchings,
expected teammate efforts must sum to total effort: every worker is
someone's teammate exactly once. To reproduce the example's interior
efforts, each low-cost worker would have to expect effort \(e_L\) from her teammate,
and each high-cost worker would have to expect effort \(\eta>e_H\).
The sum of these expected efforts exceeds the workers' total effort. No common prior reproduces the example's effort profile in a Bayesian
equilibrium without a matching signal.

At the cutoff there is a sharper comparison. Return to the random-matching
benchmark of Section~\ref{sec:benchmark-matching}: an ingroup match reveals
the teammate's identity and an outgroup match gives the signal
\(\mathrm{Out}\). Allow the designer to choose the distribution from which the matching
is drawn, keeping the signal rule fixed. For every such distribution,
equilibrium effort, averaged over workers and the draw, is bounded
above by the population average of \(1/(c_i-t)\). Under optimal OHE matching,
limiting effort equals \(1/(c-t)\) up to \(c^*\) and strictly exceeds
it above \(c^*\), a set of positive mass. Hence Bayesian play cannot
even reproduce the limiting average effort.

The comparison also holds for expected successes, which depend on whose
efforts are paired. At a fixed matching, the members of each ingroup pair best respond to one
another under both Nash and OHE, so their efforts agree exactly.
Optimizing separately, the designer can attain the Nash optimum by
pairing consecutive workers in the pooled cost ranking. This gives the limiting effort schedule \(1/(c-t)\) (Appendix~\ref{proof:org-nash-benchmark}). If the designer selects that matching with probability one, the random-matching
benchmark yields the same efforts and expected successes. Appendix~\ref{proof:org-bayesian-value} bounds expected successes under every prior and shows that the best Bayesian value per task has the same limit as the optimized Nash value. Relative to
either benchmark, the entire limiting gain in expected successes per
task under optimal OHE matching comes from above the cutoff:
\[
 \int_{(c^*,b]}\!
 \left[
 p\bigl(e^*(c),e^*(c)\bigr)
 -
 p\!\left(\frac{1}{c-t},\frac{1}{c-t}\right)
 \right]dF(c)>0 .
\]
Thus the gain over both optimized benchmarks is strict in all
sufficiently large organizations along the sequence.

Along these sequences of optimal matchings, both ingroup and outgroup teams remain present in the limit. Apart from a vanishing share of workers, those below the cutoff work ingroup and those above it work outgroup. The low-cost workers' effort contributes directly to the success of their own tasks and raises the effort that workers in the other group expect from an outgroup teammate. Above the cutoff, that average elicits more effort than a similarly costly teammate would elicit in Nash equilibrium. The choice of teams thus lets the designer preserve productive ingroup partnerships while encouraging effort on other tasks.

\section{Coordination}\label{sec:coordination-application}

Can investment by two players who are matched together induce
investment in pairs that neither belongs to? Each player's payoff
depends only on her own pair. Her choice depends on how likely she thinks her teammate is to invest.
When the teammate belongs to the other group, investment elsewhere in
that group shapes what she expects her teammate to do, even when she
knows exactly whom she faces. Decisions in different pairs can thus be
linked without any player having to reason about how others form their expectations.

There are two groups, \(G_b=\{b_1,\ldots,b_n\}\) and
\(G_d=\{d_1,\ldots,d_n\}\), with the same ordered characteristics
\(x_{b_k}=x_{d_k}=x_k\). The two actions are \emph{Invest} (\(I\)) and \emph{Not Invest} (\(NI\)).
A player who does not invest earns zero. If \(i\) invests with partner \(j\), she earns
\((x_i+x_j)/2\) when \(j\) also invests and one unit less otherwise.
Player \(i\) observes that her partner is \(j\) and assigns probability \(\beta_i(NI)\)
to \(j\) not investing. Her expected gain from investing is
\[
  \frac{x_i+x_j}{2}-\beta_i(NI).
\]
The pair's mean characteristic measures its return to joint investment.
The less likely she thinks her partner is to invest, the higher that
return must be for her to invest.

\paragraph{How investment spreads across pairs.}
Let \(n\ge2\) and \(x_k=k(1+\varepsilon)/n\), with
\(0<\varepsilon<1/(4n^2-1)\). When \(b_n\) and \(d_n\) meet, their
mean characteristic exceeds one, so each invests even if the other does not.
Every other pair has mean strictly between zero and one. Its two-player
game has a Nash equilibrium in which both players invest and another
in which neither does.

Fix a matching that pairs players by rank across groups. As in Section~\ref{section:strategic_environment}, each player observes her own partner. Under OHE, she assigns investment its average probability in that partner's group. Consider \(b_{n-1}\). She knows that
she faces \(d_{n-1}\), yet investment by \(d_n\) already contributes
\(1/n\) to the probability she assigns to investment by \(d_{n-1}\). The same reasoning
applies on the other side of the match.

\begin{proposition}\label{prop:coordination-grid}
Under the fixed matching \(\varphi(b_k)=d_k\) for every \(k\), with each player observing her own partner, full
investment is the unique OHE.
\end{proposition}

The argument works down the ranks: investment by each pair makes investment strictly worthwhile for the next, so no player randomizes in equilibrium (Appendix~\ref{proof:coordination-grid}).

\bigskip
Could uncertainty about the matching also make full investment the
unique prediction? In the random-matching benchmark of
Section~\ref{sec:benchmark-matching}, the matching is drawn according
to the commonly known distribution \(\mu\). Each player observes her partner's
identity in an ingroup match and receives only \(\mathrm{Out}\) in an
outgroup match. Conditional on her signal, she evaluates the payoff for each possible
partner using that partner's characteristic and equilibrium play. In particular, whenever \(b_n\) and \(d_n\)
meet, nobody else can meet either one. Only matchings in which the two are apart contribute to another
player's probability of meeting either of them.

\begin{proposition}\label{prop:coordination-bayesian}
For every prior \(\mu\in\Delta(\Phi)\), there is a pure BNE with full
investment and another in which some player chooses \(NI\) at a signal of
positive probability.
\end{proposition}

Full investment is therefore possible in both settings. Under any common prior, however, the Bayesian game also admits an equilibrium in which investment is not universal (Appendix~\ref{proof:coordination-bayesian}).

\paragraph{When only some pairs invest.}
Full investment illustrates one possible outcome. To allow the same
reasoning to operate from both ends, fix \(n\ge3\), take any ordered
interior list \(0<x_2\le\cdots\le x_{n-1}<1\), and let
\(x_1=-\varepsilon_L\) and \(x_n=1+\varepsilon_H\), with both excesses
positive. Under rank matching neither member of the bottom pair invests, and both members of the top pair invest. Each pair whose members must choose \(NI\) raises by \(1/n\) the minimum probability any player assigns to her partner choosing \(NI\). Each pair whose members must invest raises the corresponding lower bound for investment by the same amount. The question is whether these two arguments determine what
every intermediate pair does.

Fix any full-outgroup matching in which each player observes her own partner,
and order its pair means as
\(m_1\le\cdots\le m_n\). Under rank matching \(m_k=x_k\). Suppose no one in the first \(k\) pairs invests and everyone else does. Since exactly \(k\) players in each group do not invest, every player assigns probability \(k/n\) to her partner choosing \(NI\). If she invests, this is the probability she assigns to investing alone. For an interior cutoff it suffices to check the two
pairs on either side. For \(NI\) to be a best response in pair \(k\), investing cannot yield a positive gain, so \(m_k\le k/n\). For investment to be a best response in pair \(k+1\), its gain must be nonnegative, so \(k/n\le m_{k+1}\).
Equality allows either action.

\begin{proposition}\label{prop:coordination-cutoff}
Under any fixed full-outgroup matching in which each player observes her own partner, the OHE is unique if and
only if exactly one \(k\in\{0,\ldots,n\}\) satisfies
\[
                          m_k\le k/n\le m_{k+1},
\]
with the left inequality omitted at \(k=0\) and the right one at
\(k=n\). In the unique OHE, no one in the \(k\) pairs with the lowest means invests; everyone else invests.
\end{proposition}

Starting best responses from no investment and from full investment gives lower and upper bounds on every equilibrium profile, including mixed profiles (Appendix~\ref{proof:coordination-cutoff}).

\bigskip
In the original example \(m_r>r/n\) at every rank. If any pairs chose \(NI\), both members of the highest-ranked such pair would prefer to invest. More generally, these inequalities characterize the
matchings that yield full investment uniquely. If any inequality fails,
the largest rank at which it fails satisfies the cutoff test. A simple
sufficient condition also gives an interior cutoff under rank matching.
If all consecutive gaps except the one between ranks \(k\) and \(k+1\)
are at most \(1/n\), each pair whose choice is determined raises the lower bound on the probability assigned to that action enough to determine the next pair's choice on its side. Then only the pairs above rank \(k\) invest.

For the Bayesian comparison, the two endpoint excesses may be chosen
independently. A single bound applies to every prior over matchings.

\begin{proposition}\label{prop:coordination-bayesian-general}
For every interior list there exists \(\bar\varepsilon>0\) such that,
whenever \(0<\varepsilon_L,\varepsilon_H<\bar\varepsilon\), the
Bayesian game has at least two pure BNE for every prior
\(\mu\in\Delta(\Phi)\).
\end{proposition}

If the prior assigns positive probability to an ingroup pair, its members can coordinate on either action at their respective identity signals. For priors supported on outgroup matchings, the construction instead separates equilibria by the ranks of the players who invest (Appendix~\ref{proof:coordination-bayesian-general}).

\paragraph{Large finite populations.}
The cutoff can persist as characteristics become dense. Let \(F\) be
a continuous, strictly increasing distribution function on \([0,1]\)
with \(F(0)=0\) and \(F(1)=1\). For a possible cutoff \(c\in(0,1)\),
consider the condition
\[
    F(x)\ge x\quad (0\le x<c),\qquad
    F(x)\le x\quad (c<x\le1).
\]
Below \(c\), at least a fraction \(x\) of the population has
characteristic at most \(x\). If none of those players invests, no one among them
gains by investing under rank matching. Above \(c\), at
least a fraction \(1-x\) has characteristic at least \(x\). If that
fraction invests, each of those players is willing to invest. The inequalities allow both
forces to operate even when the distribution has positive density at
the cutoff.

\begin{corollary}\label{cor:coordination-finite}
For any \(F\) and \(c\) satisfying these conditions, there is a
sequence of symmetric finite populations whose empirical distribution functions
converge uniformly to \(F\) and whose endpoint excesses tend to zero.
At every sufficiently large group size \(n\), the unique OHE under rank matching
has the following form: no one in the lowest \(\lfloor nc\rfloor\) pairs invests,
and everyone in the remaining pairs invests. Every prior
over matchings admits at least two pure BNE. Both conclusions hold on
an open neighborhood of each such list within the symmetric populations.
\end{corollary}

Appendix~\ref{proof:coordination-finite} constructs the finite lists from quantiles of \(F\).

\bigskip
The uniform distribution satisfies the condition for every interior
\(c\). Thus finite populations approaching the same distribution can
have different, uniquely determined investment rates. In each sequence
the OHE investment rate tends to \(1-c\). For those very populations,
under any prior supported on full-outgroup matchings, the two Bayesian equilibria constructed in Appendix~\ref{proof:coordination-bayesian-general}
have investment rates at most \(1/n\) and at
least \(1-1/n\), respectively. The difference comes from whose actions shape a player's expectations.
Under OHE, a player who invests for sure contributes to the probability
that every player in the opposite group assigns to investment by her partner. In the Bayesian
game, her investment matters to another player to the extent that they
can actually meet.

\paragraph{What the cascade selects.}
The same mechanism can also yield a unique outcome in which no one invests. In both
groups, replace the original ordered list by the reflected list whose
rank \(k\) has characteristic \(1-x_{n+1-k}\). Match the two groups
outgroup by rank as before. The lowest rank has a negative
characteristic, so neither member of that pair invests. If no one at the \(k-1\) lower ranks
invests, the probability each player assigns to her partner choosing \(NI\) is at least \((k-1)/n\),
strictly above \(1-x_{n+1-k}\). Investing then gives each member of pair \(k\) a strictly
negative expected payoff, even if the remaining players mix.
Hence no one invests in the unique OHE.

Information about the payoff matrix matters in the electronic mail game
of \citet{rubinstein1989electronic}, where messages may fail to arrive.
In the OHE comparison, each player observes both characteristics in her pair. The
result also differs from the risk-dominance
selection in the global games of \citet{CarlssonVanDamme1993}, where
players receive noisy signals about the payoff matrix. In the original
grid with \(n\geq 3\), the risk-dominant equilibrium of the lowest pair
in isolation is the one in which neither player invests. Its mean characteristic is
below \(1/2\), so each player prefers not to invest against an opponent
who invests with probability \(1/2\). In the unique OHE, both members of that pair invest because investment
elsewhere raises the probability each assigns to her partner investing. The
propagation resembles the argument in \citet{Winter2004}, where a
reward makes investment dominant for one player and this in turn
makes investment attractive to others. Here investment in one pair affects what players in other pairs expect
of their partners.\footnote{\citet{SteinerStewart2008}
study the same two-player payoff matrix with an exogenous state in place
of the pair mean. Players extrapolate from past games with similar
payoffs, so a dominant action in one game can affect play in another.
Here, when a player has a dominant action, her choice enters the group
average that shapes expectations in other pairs of the same matching.}

\section{Further Modeling Discussions}
\label{section:extensions}

\subsection{Decentralized matching}
\label{sec:decentralized-matching}

Would workers choose the same teammates as the designer? Fix a
matching and its OHE effort profile. Suppose that a worker assessing a
prospective teammate from her own group observes that worker's current
effort, whether the latter works ingroup or outgroup. For a prospective
outgroup teammate she uses the current mean effort of the entire
opposite group. She assesses her incumbent teammate by the same rule
and holds her own effort fixed in both comparisons. At a given own
effort, her payoff is strictly increasing in teammate effort. She
therefore prefers whichever teammate she expects to exert more.

Call a matching \emph{pairwise stable at current efforts} if no two
workers on different tasks both strictly prefer to work with each
other. Their former teammates can be paired together; their consent is
not required. Workers compare two partners, with no option of leaving
the organization. This is a test of a proposed rematch at current
efforts, not a prediction of the efforts after it takes place.\footnote{%
\citet[Chap.~2]{goursat:tel-04757027} also studies how workers evaluate
prospective matches using observations from current matches; there the
observations are realized match utilities. For a distributional treatment
of pairwise stability in large two-sided markets, see
\citet{greinecker2021pairwise}.}

The distinction between \emph{who works outgroup} and \emph{with whom}
matters here. Suppose first that the groups have identical finite cost
lists. Call an assignment symmetric if the two groups have the same
cost pairs on their ingroup tasks and the same list of outgroup costs,
counting multiplicities. There always exists a symmetric stable
matching. As the common cost list approaches the distribution in
Proposition~\ref{prop:large-finite-cutoff}, every sequence of symmetric
stable matchings has the designer's limiting cutoff \(c^*\): apart from
a vanishing fraction in each group, workers below it work ingroup and
workers above it work outgroup. The reason is easiest to see in the limit.
Low-cost workers have ingroup teammates who exert
more than the opposite group's mean. High-cost workers exert less than
that mean, so two of them from the same group would not choose one
another in preference to their outgroup teammates. Online Appendix~\ref{proof:decentralized-cutoff} proves existence and the limiting result.\footnote{Proposition~\ref{prop:large-finite-cutoff}
also allows different finite cost lists approaching the same law. At
any sequence of designer optima on those lists, every blocking pair
contains at least one member of an exceptional set whose share tends to
zero (Online Appendix~\ref{proof:decentralized-general}). This does not assert exact stability at every finite optimum.}
Symmetry of the assignment matters:
identical populations can also admit asymmetric stable assignments (Online Appendix~\ref{oa:asymmetric-stability}).

Once the ingroup teams and outgroup workers are fixed, \emph{every}
pairing between the two outgroup lists preserves efforts and stability.
A worker already matched outgroup expects the same effort from every
prospective outgroup teammate. The designer, by contrast, cares which
efforts enter the same task: pairing high effort with high effort
raises expected success whenever it corrects a strict reversal.
Pairing outgroup workers by cost rank attains the designer's limiting
success per task along the symmetric stable sequences just described.
Workers' willingness to rematch does not select that pairing.

Why use current efforts? In an OHE, a worker chooses effort given the effort she expects
from her teammate. This rematching rule asks the worker to apply
the same distinction between ingroup and outgroup to a prospective
teammate. Even letting her choose her own effort anew for each prospective
teammate, taking expected teammate effort as given, would leave her ranking unchanged, since a higher teammate
effort raises her payoff at every own effort. A different rule would
ask her to anticipate how others change effort when teams change, and
how those changes feed back into both group means. That calls for
reasoning about responses to a counterfactual matching that the
baseline OHE does not require.

\subsection{Ingroup homogeneity and analogy partitions}
\label{sec:analogy-partitions}

In an analogy-based expectation equilibrium \citep{Jehiel2005},
a player uses the same conjecture about behavior in every decision situation
in an analogy class. Her conjecture equals the average mixed action in
the class, weighted by the frequencies of its constituent situations. In the
unperturbed finite-action game with a fixed matching, each player has one
decision situation, reached with probability one. For each observer,
place every other ingroup player in a separate class and put the whole
outgroup in one class, when nonempty. The outgroup class includes players
who are themselves matched ingroup. The player's conjecture equals her actual partner's mixed action in an ingroup match
and the average mixed action in the partner's group in an outgroup match. Under these
partitions, ABEE and OHE have the same mixed-action profiles
(Online Appendix~\ref{app:ohe-abee}).

Group membership determines which players are distinguished; matching
determines which conjecture applies to the actual partner. Even when
the partitions stay fixed, a change in matching can change the behavior
being averaged. A player whose partner stays the same may therefore
change both her conjecture about that partner and her own action. This distinction carries over to
other fixed partitions. If the same player can face different partners,
the choice of what to pool also matters: averaging that player's
actions across encounters can conceal differences in how she plays
against different partners. Online Appendix~\ref{oa:encounter-feedback} compares action frequencies by role and by encounter.

In OHE, conjectures about ingroup partners are partner-specific. Yet how similar people
perceive the members of a group to be can vary with the setting
\citep{MullenHu1989,brewer1993social}. To vary the degree of
pooling, partition a player's potential ingroup partners
into \(K_{\mathrm{In}}\) classes and the potential outgroup partners
into \(K_{\mathrm{Out}}\) classes. Her conjecture about her actual partner's action is the average action
distribution in that partner's class. For
example, she could pool outgroup members within each occupation while
distinguishing members of different occupations. The number of classes
alone does not say which distinctions she makes. Splitting an existing
class retains the previous distinctions and adds a new one.

These partitions can be specified in advance while keeping group
membership fixed. A further question is how players choose which
distinctions to make. \citet{jehiel2025endogenous} give a point of
departure by letting players choose a partition of games to minimise
prediction error for a given number of classes. Applying that approach
here would require specifying which encounters supply the observations
and whether players can allocate their classes between ingroup and
outgroup partners.

\subsection{Groups as networks}
\label{sec:groups-as-networks}

A network offers another way to specify whose behavior players distinguish.
Keep the matching as the rule that determines whom each player meets.
A separate graph can specify when a player uses her partner's actual
action distribution: she does so if they are linked. If they are not
linked, the model must also specify whose actions she pools. One rule is to
average the action distributions of all other players to whom she
is not linked. If all players in the same group are linked
and there are no links between groups, this rule recovers OHE.
On other graphs, members of the same group may pool the actions of
different players when forming their conjectures.

The distinction between the matching and the graph is useful for
comparing OHE with the peer-confirming equilibrium of
\citet{lipnowski2019peer}. Along each link in their graph, a player's conjecture about the
other player's strategy is correct. Common belief of rationality and of
correct conjectures along links also restricts what players can believe
about those to whom they are not linked. In OHE, the link pattern
similarly tells a player when to use a partner-specific conjecture;
for an outgroup partner, the pooling rule determines her conjecture.

In the network games of \citet{BattigalliPanebiancoPin2023}, links
determine how neighbors' actions enter a player's payoff. Players
learn from their realized payoffs. When externalities are only
local, an active player can infer the aggregate effect of her
neighbors' actions, without necessarily identifying their
individual actions. A payoff at zero activity does not reveal
that aggregate. In the learning interpretation of OHE
(Section~\ref{subsec:learning-foundation_ABEE}), a subject observes
the preceding period's action frequencies pooled across the opposite
group. When she meets an outgroup partner, her payoff depends on that
partner's action, but the choices of other group members also shape what
she expects that partner to do.

\section{Related Literature}
\label{sec:related-literature}

The outgroup homogeneity effect concerns how much variation people
perceive among members of their own and other groups
\citep{MullenHu1989}. Its role here is specific: a player may know
exactly how two outgroup members differ in characteristics and yet
have the same expectations about how either one would act. The group label
need not enter her utility, as it can in models of identity
\citep{AkerlofKranton2000}. The object of her assessment also differs
from those in other accounts of group differences.
In \citet{LiquiLung2022}, representation of one's own group among
successful people helps guide choices when self-assessments of ability
are noisy. In \citet{ChauvinWP}, people attribute differences in
outcomes across groups to differences in traits. In OHE, the behavior of the group shapes what a player expects her
outgroup teammate to do.

The closest formal connection is to analogy-based expectation
equilibrium \citep{Jehiel2005}. Section~\ref{sec:analogy-partitions}
gives the partitions for which the two notions yield the same mixed-action profiles
in the unperturbed finite-action game at a fixed matching. What matching
adds can be seen by keeping a player's outgroup teammate fixed and changing
who works together elsewhere. If those changes affect how members of
her teammate's group act, they can change what she expects that same teammate to do.\footnote{Other models place the coarse assessment on
the player's own alternatives: \citet{OsborneRubinstein1998sampling}
compare the consequences sampled from each own action, and
\citet{JehielSamet2007valuation} assign a common valuation to similar
moves. Here each player maximizes expected utility given her conjecture.}
A related distinction arises in cursed equilibrium
\citep{EysterRabin2005}. A fully cursed player correctly anticipates
how often opponents take each action but neglects how their choices
depend on their private information. \citet{JehielKoessler2008}
show how their private-information analogy partition yields fully
cursed play in a two-player Bayesian game. In OHE the partner's identity and payoff-relevant
characteristics are observable. The averaging is across different
people, including outgroup members whom the player cannot meet at the
given matching.

Self-confirming equilibrium separates optimal choice from what the
player's observations can establish about her conjecture
\citep{battigalli1987comportamento,FudenbergLevine1993SCE}. In a
static game with feedback, a player best responds to a conjecture
that correctly predicts the distribution of feedback at the action she takes
\citep[Sec.~6.3]{BattigalliCatoniniDeVito2025}. What that condition
requires depends on the feedback. In the investment game of
Section~\ref{sec:coordination-application}, a player who does not invest
earns zero regardless of her teammate's action. Observing this payoff
does not tell her whether the teammate invested. Observing the
teammate's action directly would distinguish the two outcomes;
\citet[Example~2]{BattigalliEtAl2016Feedback} give an example of
such action feedback.

In the learning model of Section~\ref{subsec:learning-foundation_ABEE},
a successor instead observes action frequencies from the preceding cohort.
For an outgroup teammate, those frequencies pool the choices of the
teammate's entire group, including its members on ingroup teams. If one
third of that group invested, the frequency in her teammate's role need
not have been one third. She nevertheless assigns probability one third to investment by her
current teammate.
\citet{HuckJehielRutter2011} make a closely related distinction in an
experiment: the report pools opponents' actions across two games,
and different game-specific frequencies can fit the same report.
Here the frequencies pool choices across people, and team formation can
change those choices.\footnote{A Berk--Nash equilibrium restricts
beliefs to the best-fitting specifications in the player's subjective
model \citep{EspondaPouza2016}. In \citet{Spiegler2016}, a subjective
causal graph determines which statistical dependencies the decision
maker retains. In OHE the pooling rule specifies which players' actions enter an
outgroup conjecture.}

The organizational application connects to work on team composition
and shared cognitive frames. \citet{Prat2002} asks whom an organization
should hire when workers' backgrounds determine the signals they
receive. When jobs are complementary, greater correlation in workers' errors
can make hiring similar types desirable. \citet{GibbonsLiCalziWarglien2021}
study how a shared way of categorizing situations affects collective
performance. In Section~\ref{section:orgs}, the workforce and group
memberships are fixed. Assigning two productive workers to the same
ingroup task affects more than their joint output: their effort
also affects what workers in the other group expect from an outgroup
teammate. This gives the
designer a reason to keep some teams within groups even when other
teams work across them.

\section{Conclusion}\label{sec:conclusion}

In an OHE, a player bases her expectations about an outgroup partner on the behavior of the partner's entire group, including its members who are matched ingroup. Group membership is fixed, but the behavior the player draws on depends on who is paired with whom.

In the organizational model, low-cost workers paired ingroup elicit high effort from one another. Their effort raises the group's average effort and hence the effort that workers in the other group expect from an outgroup teammate. In the large organizations studied here, as the two groups approach the same cost distribution, every matching that maximizes expected task successes follows a common cutoff apart from a vanishing share of workers: those below it work ingroup and those above it work outgroup.

In the coordination example, the highest-ranked players in the two groups invest together. A player in another pair takes their investment into account when forming expectations about her own partner. Full investment is the unique OHE under rank matching. In the random-matching benchmark, every common prior admits both a pure equilibrium with full investment and another in which a player does not invest at a signal she receives with positive probability.

%--------------%
%-- APPENDIX --%
%--------------%

\clearpage
\appendix
\pdfbookmark[0]{Appendix}{main-appendix}
\section*{Appendix}
\section{Proofs of the main results}\label{app:proofs}
The learning result is proved separately in Appendix~\ref{proof:learning-foundation}.

\subsection{Existence of an Outgroup Homogeneity Equilibrium}\label{proof:ohe-existence}
\noindent\textit{Proof of Proposition~\ref{prop:ohe-existence}.} We apply Kakutani's fixed-point theorem to the correspondence of
best responses to consistent conjectures.

Fix \(x\), \(\varphi\), and \((G_1,G_2)\).
For each profile \(\sigma\in(\Delta(A))^I\), let
\(\beta_i(\sigma)\) be the conjecture prescribed by the consistency
rule, and define
\[
\mathcal R_i^\varphi(\sigma)
=
\operatorname*{arg\,max}_{\sigma_i'\in\Delta(A)}
U_i(\sigma_i',\beta_i(\sigma);x_i,x_{\varphi(i)}),
\quad
\mathcal R^\varphi(\sigma)
=
\prod_{i\in I}\mathcal R_i^\varphi(\sigma).
\]
The conjecture \(\beta_i(\sigma)\) is held fixed during maximization.

The domain \((\Delta(A))^I\) is a nonempty, compact, convex subset
of \(\mathbb R^{I\times A}\). Continuity and affinity of \(U_i\)
in its first argument ensure that each
\(\mathcal R_i^\varphi(\sigma)\) is nonempty and convex.
Thus \(\mathcal R^\varphi\) maps this domain into itself and has
nonempty convex values.

Each conjecture map is continuous, being either a coordinate
projection or a fixed finite average over a nonempty group.
To verify that \(\mathcal R^\varphi\) has a closed graph, take
\(\sigma^n\to\sigma\) and
\(\widehat\sigma^n\to\widehat\sigma\) in \((\Delta(A))^I\),
with \(\widehat\sigma^n\in\mathcal R^\varphi(\sigma^n)\).
For every \(i\in I\) and \(\sigma_i'\in\Delta(A)\),
\[
U_i(\widehat\sigma_i^n,\beta_i(\sigma^n);x_i,x_{\varphi(i)})
\ge
U_i(\sigma_i',\beta_i(\sigma^n);x_i,x_{\varphi(i)}).
\]
Since \(\beta_i(\sigma^n)\to\beta_i(\sigma)\) and \(U_i\) is a
finite sum of products of mixed-action probabilities and fixed
real payoffs, taking limits gives
\(
U_i(\widehat\sigma_i,\beta_i(\sigma);x_i,x_{\varphi(i)})
\ge
U_i(\sigma_i',\beta_i(\sigma);x_i,x_{\varphi(i)})
\). 
As \(i\) and \(\sigma_i'\) were arbitrary,
\(\widehat\sigma\in\mathcal R^\varphi(\sigma)\), proving that
the graph is closed.

Kakutani's fixed-point theorem therefore ensures the existence of a fixed point
\(\sigma^*\in\mathcal R^\varphi(\sigma^*)\).
Set \(\beta_i^*=\beta_i(\sigma^*)\) for every \(i\).
This conjecture profile is consistent with \(\sigma^*\), and the
fixed-point condition ensures that every \(\sigma_i^*\) is a best
response to \(\beta_i^*\). Hence \(\sigma^*\) is an OHE under
\(\varphi\). \(\blacksquare\)

\subsection{Equivalence under uniformly random outgroup matching}\label{proof:random-equivalence}
\noindent\textit{Proof of Proposition~\ref{prop:ohe-bne-equivalence}.} Fix \(x\), \(\varphi\in\Phi_{\mathrm{Out}}\), and \(\sigma\). Since $\mu$ is the uniform distribution over $\Phi_{\mathrm{Out}}$, every player receives the signal \(\mathrm{Out}\).
Let \(s\) be the profile of decision functions defined by \(s_j(\mathrm{Out})=\sigma_j\) for every \(j\in I\). For every \(i\in I\) and \(\sigma_i'\in\Delta(A)\),
\[
\begin{aligned}
\bar U_i(\sigma_i',s_{-i};\mathrm{Out})
&=
\sum_{\psi\in\Phi_{\mathrm{Out}}}
\mu(\psi\mid\mathrm{Out})\,
U_i(\sigma_i',s_{\psi(i)}(\mathrm{Out});x_i,x_{\psi(i)})\\
&=\frac1n\sum_{j\in I\setminus G(i)}
U_i(\sigma_i',\sigma_j;x_i,x_j)=U_i\left(
\sigma_i',\bar\sigma_{I\setminus G(i)};
x_i,x_{\varphi(i)}
\right),
\end{aligned}
\]
where the second equality holds because $\mu$ is uniform on \(\Phi_{\mathrm{Out}}\) and \(s_j(\mathrm{Out})=\sigma_j\) for every $j$, while the last equality uses private values and affinity in the partner's mixed action. Under \(\varphi\), consistency uniquely determines the conjectures \(\beta_i=\bar\sigma_{I\setminus G(i)}\) for every
\(i\). With these conjectures, the identity makes Bayesian and OHE
optimality equivalent for every own mixed deviation, proving both
directions.
\(\blacksquare\)

\subsection{Existence and uniqueness of equilibrium effort}\label{proof:org-ohe}
\noindent\textit{Proof of Proposition~\ref{prop:org-ohe}.} Fix the characteristic profile, the groups, and the matching
\(\varphi\). For a conjecture \(\beta_i\), let
\(m_i=\int_{[0,\bar e]}s\,\beta_i(ds)\in[0,\bar e]\). The payoff from a pure
effort \(e_i\) is
\[
U_i(\delta_{e_i},\beta_i;x_i,x_{\varphi(i)})
=\kappa\bigl[e_i+(1+te_i)m_i\bigr]-\frac{x_i e_i^2}{2}.
\]
Replacing \(\beta_i\) with \(\delta_{m_i}\) leaves this expression unchanged
for every pure effort and hence for every own mixed action.
Its derivative is \(\kappa(1+tm_i)-x_i e_i\), and its second
derivative is \(-x_i<0\). Since \(c_i=x_i/\kappa>0\), its unique
maximizer on \([0,\bar e]\) is
\[
\min\left\{\bar e,\frac{1+tm_i}{c_i}\right\}.
\]
Any mixed action assigning positive probability to other efforts
has strictly smaller expected payoff. An optimal mixed action must
therefore assign probability one to this unique maximizer. Thus
every OHE is pure, including when the cap binds or
the unconstrained maximizer equals the cap.

For \(e\in[0,\bar e]^I\), let \(m_i(e)\) be as in~\eqref{eq:org-means-map}.
Define \(T:[0,\bar e]^I\to[0,\bar e]^I\) by
\[
T_i(e)=\min\left\{\bar e,
\frac{1+t m_i(e)}{c_i}\right\}.
\]
The map \(T\) is continuous: each \(m_i(e)\) is a coordinate or
a finite arithmetic average, and the minimum of two continuous
functions is continuous. Its coordinates lie in \([0,\bar e]\).
Since this cube is nonempty, compact, convex, and finite-dimensional,
Brouwer's fixed-point theorem gives a fixed point \(e\).

At any such fixed point, set \(\sigma_i=\delta_{e_i}\) and
\[
\beta_i=
\begin{cases}
\delta_{e_{\varphi(i)}},
& i\in\operatorname{In}_{G(i)}^\varphi,\\[1mm]
\displaystyle\frac{1}{|G(\varphi(i))|}
\sum_{j\in G(\varphi(i))}\delta_{e_j},
& i\in\operatorname{Out}_{G(i)}^\varphi.
\end{cases}
\]
These conjectures satisfy distributional consistency, and their
means are \(m_i(e)\). The fixed-point equations therefore
make each \(\sigma_i\) a best response to \(\beta_i\), including
against mixed deviations with the conjecture held fixed. Hence
\(\sigma\) is an OHE. Conversely, purity and consistency imply
that every OHE arises from a fixed point of \(T\).

To establish uniqueness, let \(e\) and \(\tilde e\) be fixed points.
Every coordinate is strictly positive, since
\(T_i(\cdot)\ge\min\{\bar e,1/c_i\}>0\). If \(e\ne\tilde e\),
interchange the profiles if necessary so that
\[
\lambda=\max_{j\in I}\frac{e_j}{\tilde e_j}>1,
\]
and choose \(i\) attaining this maximum. Then
\(\tilde e_i=e_i/\lambda<\bar e\), so the cap is slack at \(\tilde e_i\) and
\(c_i \tilde e_i=1+t m_i(\tilde e)\). Moreover,
\(e_j\le\lambda \tilde e_j\) for every \(j\) implies
\(m_i(e)\le\lambda m_i(\tilde e)\). Consequently,
\[
e_i
\le\frac{1+t m_i(e)}{c_i}
\le\frac{1+t\lambda m_i(\tilde e)}{c_i}
=\lambda \tilde e_i-\frac{\lambda-1}{c_i}
<\lambda \tilde e_i=e_i,
\]
a contradiction. The effort fixed point, and therefore the OHE,
is unique. \(\blacksquare\)

\subsubsection{Interior efforts and group means}\label{proof:org-interiority}
\noindent\textit{For a fixed matching, condition~\eqref{eq:org-interiority} ensures an interior best response at every admissible mean. Equilibrium efforts are then given by~\eqref{eq:org-efforts}, and the means of two nonempty groups uniquely solve~\eqref{eq:org-means}.}

For a fixed agent, the unconstrained response
\((1+tm_i)/c_i\) is positive and increasing in
\(m_i\in[0,\bar e]\). It is strictly below \(\bar e\) for
every admissible mean if and only if
\[
\frac{1+t\bar e}{c_i}<\bar e
\quad\Longleftrightarrow\quad
c_i>t+\frac{1}{\bar e}.
\]
Thus condition~\eqref{eq:org-interiority} is necessary and
sufficient for uniform strict response interiority. At equality,
\(c_i=t+1/\bar e\), the response reaches the cap exactly when
\(m_i=\bar e\); it is strictly interior for every smaller mean.
If \(c_i<t+1/\bar e\), the unconstrained response at
\(m_i=\bar e\) exceeds the cap. Interiority at the particular
equilibrium requires only
\(1+t m_i(e(\varphi))<c_i\bar e\) for each \(i\).

Suppose henceforth that~\eqref{eq:org-interiority} holds.
For an ingroup pair \(j=\varphi(i)\), the equilibrium equations are
\[
c_i e_i(\varphi)=1+t e_j(\varphi),
\qquad
c_j e_j(\varphi)=1+t e_i(\varphi).
\]
Since \(c_i,c_j>t\), their determinant \(c_ic_j-t^2\) is
positive. Solving gives
\[
e_i(\varphi)=\frac{c_j+t}{c_ic_j-t^2},
\qquad
e_j(\varphi)=\frac{c_i+t}{c_ic_j-t^2}.
\]
For an outgroup worker \(i\in G\), the effort she expects from her teammate
equals the opposite group's mean by consistency. Hence
\[
e_i(\varphi)=\frac{1+t\eta_{G'}(\varphi)}{c_i},
\qquad G'=G(\varphi(i)).
\]
These are the two expressions in~\eqref{eq:org-efforts}.

When both groups are nonempty, summing these efforts over the
agents in each group gives~\eqref{eq:org-means}:
\[
|G|\eta_G(\varphi)
=\sum_{i\in\operatorname{In}_G^\varphi}e_i(\varphi)
+\bigl(1+t\eta_{G'}(\varphi)\bigr)
\sum_{i\in\operatorname{Out}_G^\varphi}\frac1{c_i},
\qquad G'\ne G.
\]
The ingroup sum counts each agent's effort, including both members
of each ingroup pair. For every nonempty group \(G\),
\[
t\sum_{i\in\operatorname{Out}_G^\varphi}\frac1{c_i}<|G|.
\]
If the outgroup subset is empty, this follows from \(0<|G|\).
Otherwise, each summand satisfies \(t/c_i<1\), and
\(|\operatorname{Out}_G^\varphi|\le|G|\). The coefficient determinant
of the two mean equations is therefore positive:
\[
|G_1||G_2|
-t^2\left(\sum_{i\in\operatorname{Out}_{G_1}^\varphi}\frac1{c_i}\right)
\left(\sum_{i\in\operatorname{Out}_{G_2}^\varphi}\frac1{c_i}\right)>0.
\]
Their unique solution is
\[
\eta_G(\varphi)=
\frac{
\begin{aligned}
&|G'|\left(
\sum_{i\in\operatorname{In}_G^\varphi}e_i(\varphi)
+\sum_{i\in\operatorname{Out}_G^\varphi}\frac1{c_i}\right)\\
&\quad+t\left(\sum_{i\in\operatorname{Out}_G^\varphi}\frac1{c_i}\right)
\left(
\sum_{i\in\operatorname{In}_{G'}^\varphi}e_i(\varphi)
+\sum_{i\in\operatorname{Out}_{G'}^\varphi}\frac1{c_i}\right)
\end{aligned}
}{
|G||G'|
-t^2\left(\sum_{i\in\operatorname{Out}_G^\varphi}\frac1{c_i}\right)
\left(\sum_{i\in\operatorname{Out}_{G'}^\varphi}\frac1{c_i}\right)
},\qquad G'\ne G,
\]
where the ingroup efforts are given explicitly by the first
expression in~\eqref{eq:org-efforts}, and empty sums are zero.
The means of the equilibrium already proved to exist solve these
equations. Nonsingularity therefore identifies the displayed
solution with those means, which belong to \((0,\bar e)\).
The associated individual expressions likewise equal the actual
equilibrium efforts and belong to \((0,\bar e)\).

If one labelled group is empty, every pair is ingroup, so the
pairwise formula applies to every agent and no empty-group mean
is defined. If both groups are nonempty but no agent works
outgroup, the two mean equations decouple, with zero outgroup
sums. This completes the characterization. \(\square\)

\paragraph{Properties of efforts and group averages.}
Fix \(\varphi\) and maintain \eqref{eq:org-interiority}. A product of nonnegative inverse powers of positive costs is convex: its logarithm is a nonnegative linear combination of the convex functions \(-\log c_i\), and exponentiation preserves convexity. Such a product is strictly convex in any collection of coordinates whose exponents are all positive.

For an ingroup pair \(j=\varphi(i)\), the effort formula in \eqref{eq:org-efforts} has the expansion
\[
\frac{c_j+t}{c_ic_j-t^2}
=\sum_{k=0}^{\infty}
\left(\frac{t^{2k}}{c_i^{k+1}c_j^k}
+\frac{t^{2k+1}}{c_i^{k+1}c_j^{k+1}}\right),
\]
which converges because \(t^2<c_ic_j\). Every summand is nonincreasing and jointly convex in \((c_i,c_j)\). The terms \(1/c_i\) and \(t/(c_ic_j)\) establish strict decrease in each cost, while \(t/(c_ic_j)\) is jointly strictly convex. The remaining sum is a pointwise limit of convex partial sums and is finite on the domain under consideration, so it is convex. Thus the ingroup effort is jointly strictly convex in the two costs. Conditional on the opposing group's mean, the outgroup effort \((1+t\eta_{G'})/c_i\) is strictly decreasing and strictly convex in \(c_i\), and affine and strictly increasing in \(\eta_{G'}\).

To establish the equilibrium properties of the group means in \eqref{eq:org-means}, define
\[
w_{ij}=\begin{cases}
1,&i\in\operatorname{In}_{G(i)}^\varphi,\ j=\varphi(i),\\[2pt]
\displaystyle\frac{1}{|G(\varphi(i))|},
&i\in\operatorname{Out}_{G(i)}^\varphi,\ j\in G(\varphi(i)),\\[3pt]
0,&\text{otherwise}.
\end{cases}
\]
These weights are nonnegative and satisfy \(\sum_jw_{ij}=1\). The equilibrium equations are therefore
\(e_i(\varphi)=c_i^{-1}+t c_i^{-1}\sum_jw_{ij}e_j(\varphi)\).
Repeated substitution gives
\begin{equation}\label{eq:org-path-series}
e_i(\varphi)
=\sum_{k=0}^{\infty}t^k
\sum_{i_1,\ldots,i_k\in I}
\frac{\displaystyle\prod_{\ell=0}^{k-1}w_{i_\ell i_{\ell+1}}}
{\displaystyle\prod_{\ell=0}^{k}c_{i_\ell}},
\qquad i_0=i,
\end{equation}
where the \(k=0\) term is \(1/c_i\). For \(K\ge0\), after retaining the terms through \(k=K\), the remainder is nonnegative and at most
\[
\bar e\left(\max_{j\in I}\frac{t}{c_j}\right)^{K+1}.
\]
This bound follows because the products of weights over all paths of a given length sum to one, each factor \(t/c_j\) is at most \(\max_j t/c_j<1\), and every terminal effort is at most \(\bar e\). The bound tends to zero and proves \eqref{eq:org-path-series}, whose first three terms are displayed in \eqref{eq:org-series}.

At fixed \(t\), each term of \eqref{eq:org-path-series} is nonincreasing and jointly convex in the cost vector. Pointwise limits that are finite preserve both properties, as does averaging over a nonempty group. Hence every effort and every group mean is nonincreasing in every cost and jointly convex in costs. For \(j\in G\), the term \(1/(|G|c_j)\) makes \(\eta_G(\varphi)\) strictly decreasing in \(c_j\). If there is an outgroup task, each group has an outgroup worker. Writing \(G'\) for the other group, for such a worker \(i\in G\), the \(k=1\) term contains \(t/(|G'|c_ic_j)\) for every \(j\in G'\). Consequently, \(\eta_G(\varphi)\) is then strictly decreasing in every cost, including those in the other group. If all tasks are ingroup, each group's mean is independent of the other group's costs.

Finally, with costs fixed, every coefficient of the power series \eqref{eq:org-path-series} in \(t\) is strictly positive. Its positive linear and quadratic terms establish strict increase and strict convexity in \(t\); the remaining terms are nondecreasing and convex. These properties hold on the interval where \eqref{eq:org-interiority} is maintained and pass to every nonempty group mean.
\(\square\)

\subsection{Optimal pairing for fixed ingroup and outgroup subsets}\label{proof:org-conditional-sorting}
\noindent\textit{Proof of Proposition~\ref{prop:org-conditional-sorting}.} An optimum exists because the feasible set is finite and nonempty.
Fix one.

For an ingroup team, let \(F(c,d)\) denote either its equilibrium success
probability or its total equilibrium effort. Each function is symmetric
and strictly supermodular: its inverse-cost expansion has nonnegative
coefficients and a positive coefficient on \(1/(cd)\). If an ingroup
pairing is not assortative, choose a cost-ordered list with the longest
initial segment consisting of consecutive matched pairs. The first two
remaining workers have costs \(a\le b\) and respective partners with costs
\(d>b\) and \(c>a\); otherwise reordering ties would extend that segment.
Thus
\[
F(a,b)+F(c,d)>F(a,d)+F(b,c).
\]
Replacing the pairs \((a,d)\) and \((b,c)\) with \((a,b)\) and \((c,d)\)
strictly raises both direct success and total ingroup effort. Other
ingroup teams are unchanged. If outgroup tasks exist,
\eqref{eq:org-efforts}--\eqref{eq:org-means} imply that both group averages
and all outgroup efforts weakly increase. This contradicts optimality.

Reassigning only outgroup partners leaves conjectured means unchanged at
every effort profile, hence preserves the equilibrium equations and their
unique solution.\footnote{This invariance does not require interiority
and remains valid when effort caps bind.}
Within each outgroup subset, equilibrium effort decreases strictly with
cost. Strict supermodularity of \(p\) therefore rules out every strict
cost inversion: exchanging the partners would raise output without
changing efforts. The outgroup pairing is consequently assortative.

Finally, all matchings satisfying both assortativity conditions differ
only through exchanges of equal-cost workers within the fixed subsets.
They therefore yield the same objective value. Since every optimum
satisfies both conditions, every matching satisfying them is optimal. $\blacksquare$

\subsubsection{Supermodularity of task success}\label{proof:org-supermodularity}
\noindent\textit{For a fixed matching, expected total task success is supermodular in the effort profile. Under~\eqref{eq:org-interiority}, its equilibrium value is supermodular in the cost profile.}

Fix the matching, groups, \(\bar e\), \(t\), and \(\kappa\).
Before equilibrium substitution, expected total task success is
\[
S(\varphi;e)=\kappa\sum_i e_i
+\frac{\kappa t}{2}\sum_i e_i e_{\varphi(i)}.
\]
Its mixed effort derivative for distinct workers is \(\kappa t\) if
they are teammates and zero otherwise. Hence it is supermodular on
the entire feasible effort cube.

For the cost comparisons, maintain \eqref{eq:org-interiority}. Every
finite partial sum of \eqref{eq:org-path-series} is a polynomial in
inverse costs with nonnegative coefficients. For nonnegative integer
exponents \(n_\ell\) and \(j\ne k\),
\[
\partial_{jk}\prod_\ell c_\ell^{-n_\ell}
=\frac{n_jn_k}{c_jc_k}\prod_\ell c_\ell^{-n_\ell}\ge0.
\]
Thus the partial sums are supermodular. Substituting these partial
sums for efforts in the objective again gives polynomials of this
form. They converge pointwise to the objective evaluated at equilibrium,
so their four-value supermodularity inequalities pass to the limit.
This proves supermodularity in costs without differentiating an
infinite series. \(\square\)

\subsection{Opening two ingroup tasks when one group mean falls}\label{proof:opening-bound}
\noindent\textit{Proof of Proposition~\ref{prop:rematch-falling-average}.} For the rematch in the proposition, write \(m_k=\eta_{G_k}(\phi)\), \(m'_k=\eta_{G_k}(\psi)\), and \(\delta_k=m'_k-m_k\). In each group \(G_k\), the original ingroup teammates are ordered so that \(c_{i_k}\le c_{j_k}\).

\paragraph{The direct lower bound.}
Write $x_k=e_{i_k}(\phi)$, $y_k=e_{j_k}(\phi)$,
$g_k=\sqrt{x_ky_k}$, $l=\min(m_1,m_2)$, $r=\max(m_1,m_2)$,
$q=\sqrt{lr}$ and $\lambda_k=t^2/(c_{i_k}c_{j_k})\in(0,1)$.
Keep the original pairs together and initially let each cheaper worker expect effort $l$ from her teammate and each dearer worker expect effort $r$.
More generally, an original pair of costs $c,d$ and efforts $x,y$ satisfies
\[
 \frac{p((1+t\alpha)/c,(1+t\beta)/d)-p(x,y)}{\kappa t}
 =(1-\lambda)(x\alpha+y\beta-2xy)+\lambda(\alpha\beta-xy),
 \qquad \lambda=t^2/(cd).
\]
This follows by expansion using $cx=1+ty$ and $dy=1+tx$.
At $(\alpha,\beta)=(l,r)$ the right-hand side for pair $k$ is
\[
 2g_k(q-g_k)+\lambda_k(q-g_k)^2
        +(1-\lambda_k)(x_kl+y_kr-2g_kq).
\]
The last term is nonnegative by the arithmetic--geometric mean inequality.
Switch next to the two proposed outgroup pairs and give each worker the
original mean of her teammate's group. The change in the sum of products
in this step is
\[
 (1+tm_1)(1+tm_2)
  (1/c_{i_1}-1/c_{j_2})(1/c_{i_2}-1/c_{j_1}).
\]
The change in the sum of efforts divided by $t$ is
$(m_2-m_1)(1/c_{i_1}-1/c_{j_2})$ if $m_2\ge m_1$, and
$(m_1-m_2)(1/c_{i_2}-1/c_{j_1})$ otherwise. Overlap makes both
comparisons nonnegative. Dropping this latter change and the two nonnegative
remainders per pair gives exactly the displayed lower bound
$\underline{\Delta S}_{\mathrm{direct}}(\phi)$ in the text on the true change in success
at the old means.

This lower bound is strict unless all four original efforts and both means
coincide. Indeed, if all pair remainders vanish, then $g_k=q$ and
$x_kl=y_kr$ for both $k$, hence $x_k=r$ and $y_k=l$ for both pairs.
If $r>l$, both groups have the same two distinct costs
$(1+tl)/r<(1+tr)/l$. The change in the sum of efforts in the preceding
step is then strictly positive. If $r=l$, all six effort levels coincide.

\paragraph{Adjustment of group means.}
Let $D$ be the actual change in success on the two selected tasks at the
original group means. Every other task has the same success as under
$\phi$ at these means. Under the proposition's premises $\delta_1<0$
rules out the equality case above, so $D>\underline{\Delta S}_{\mathrm{direct}}(\phi)$.
First move $m_2$ to $m'_2$ holding $m_1$ fixed, and then move $m_1$ to
$m'_1$ holding $m'_2$ fixed. The task success function is affine in each
effort separately. Thus
\[
 \frac{S(\psi)-S(\phi)}{\kappa t}
 =\frac{D}{\kappa t}
 +\delta_2\sum_{i\in\mathrm{Out}^{\psi}_{G_1}}
  \frac{1}{c_i}\left(1+t\frac{1+tm_1}{c_{\psi(i)}}\right)
 +\delta_1\sum_{j\in\mathrm{Out}^{\psi}_{G_2}}
  \frac{1+t e_{\psi(j)}(\psi)}{c_j}.
\]
Each sum counts each new outgroup task once through the worker whose effort
changes in that step. The last coefficient is positive because the rematch
creates two outgroup tasks. Substituting the strict bound for $D$ and
rearranging proves the stated sufficient condition, including strict
improvement at equality in the bound. Reversing group labels gives the
other orientation. \(\square\)

\subsection{Consecutive ranks of ingroup teammates}\label{proof:ingroup-adjacency}
\noindent\textit{Proof of Proposition~\ref{prop:consecutive-ingroup-ranks}.} Let \(D\) be the group containing \(i,j,k\) and \(B\) the other group.
Write \(a=1/c_i\), \(y=1/c_k\), \(b=1/c_j\) and
\(w=1/c_{\phi(k)}\), so \(a>y>b\). For
\(z\in\{b,y,a\}\), let \(S_z\) be expected successes when the worker
with reciprocal cost \(z\) works with \(\phi(k)\), the other two work
together and all other partnerships remain fixed. All efforts in
\(S_z\) are equilibrium efforts under the indicated matching. Thus
\(S_y=S(\phi)\), \(S_a=S(\psi_i)\), and \(S_b=S(\psi_j)\).

An interpolation of these three values will give the strict inequality.
Set \(T=a+y+b\), \(J=ayb\), \(d=t^2J\), and \(s=z-d\).
Condition~\eqref{eq:org-interiority} gives \(ta,ty,tb<1\), so \(d<b\) and \(s>0\)
on \([b,a]\). At each of the three assignments the two ingroup teammates'
reciprocal costs \(u,v\) satisfy \(u+v=T-z\) and \(uv=J/z\).
Their efforts are \(u(1+tv)/(1-t^2uv)\) and
\(v(1+tu)/(1-t^2uv)\). Substitution gives the following interpolations
of their effort sum and effort product:
\[
\begin{aligned}
 E(z)&=T-z-d+\frac{tJ(2+tT-td)}{s},\\
 P(z)&=J\left[-t+\frac{1+tT-2td}{s}
                 +\frac{d(2+tT-td)}{s^2}\right].
\end{aligned}
\]
Since \(T>3d\), both inverse-power coefficients in \(P\) are positive.

Write \(N_G=|G|\) and \(h_G=1+t\eta_G\). In \(B\) the sum of
ingroup efforts \(L_B\) and the sum of outgroup workers' reciprocal
costs \(A_B\) are fixed. Its mean equation gives
\(h_B=K+\beta h_D\), where
\(K=1+tL_B/N_B\ge 1\) and \(\beta=tA_B/N_B\).
In \(D\), let \(L_0\) be the sum of ingroup efforts outside the triple
and \(A_0\) the sum of reciprocal costs of outgroup workers outside the
triple. Substitution in the other mean equation gives
\[
 h_D(z)=
 \frac{N_D+tL_0+tE(z)+tK(A_0+z)}
      {N_D-t\beta(A_0+z)} .
\]
The denominator is positive at \(a\) because \(\beta<1\) and
\(t(A_0+a)<N_D\), and it weakly decreases with \(z\).
Its reciprocal thus has a convergent power series in \(s\) with
nonnegative coefficients on \([b-d,a-d]\).
The numerator takes the form
\[
 N_D+tL_0+t(T-2d)+tK(A_0+d)
   +t(K-1)s+\frac{t^2J(2+tT-td)}{s}.
\]
All its coefficients are nonnegative. The same holds for the resulting
expansions of \(h_D\) and \(h_B=K+\beta h_D\) in integer powers of \(s\).

Let \(Q_0\) sum the products of teammates' reciprocal costs on the
unchanged outgroup tasks. At the three assignments, total equilibrium
success divided by \(\kappa\) equals the following expression up to a
constant independent of \(z\):
\[
 \frac{N_D h_D(z)+N_B h_B(z)}{t}
   +t\left[P(z)+(Q_0+wz)h_B(z)h_D(z)\right].
\]
The first term accounts for total effort. The bracket contains the
variable part of the sum of the products on actual tasks.
Since \(Q_0+wz=Q_0+wd+ws\), this expression is a constant plus a
convergent sum of integer powers of \(s\) with nonnegative coefficients.
Every such power is convex on \(s>0\), and the coefficient on \(s^{-2}\)
in \(tP(z)\) is strictly positive. The interpolation is therefore strictly
convex in \(z\), which gives
\[
 S_y<\frac{a-y}{a-b}S_b+\frac{y-b}{a-b}S_a
     \leq \max\{S_b,S_a\}.
\]

Finally, the strict two-task improvements in Appendix~\ref{proof:org-conditional-sorting}
give an ordering of the ingroup workers by cost in which teammates are
consecutive at any matching with no such improvement. For each outgroup
cost \(x\), insert the workers of that cost after the last complete pair
whose upper cost is at most \(x\), or before the first pair if there is
no such pair. If a next pair exists, its lower cost must be at least
\(x\), by the strict inequality just proved. Inserting the outgroup
workers in cost order therefore preserves both nondecreasing costs and
adjacency of every ingroup pair. If the ingroup subset is empty the
conclusion is immediate. All the exchanges used here preserve the number
of outgroup tasks, so the argument also applies at an optimum conditional
on any feasible such number.\hfill\(\square\)

\subsection{The common cost cutoff in large organizations}\label{proof:large-finite-cutoff}
\noindent\textit{Proof of Proposition~\ref{prop:large-finite-cutoff}.} Each group has \(n\) workers. The empirical cost laws converge weakly to the same absolutely continuous \(F\) on \([a,b]\), where \(a>t+1/\bar e\). The effort cap and task technology are fixed as in Section~\ref{section:orgs}, with \(0<t<1\) and \(0<\kappa\le(2\bar e+\bar e^2)^{-1}\). Write \(S_n^*=\max_\phi S(\phi)\) for the largest expected number of successful tasks in organization \(n\). The proof also covers any sequence \(\phi_n\) with \((S_n^*-S(\phi_n))/n\to0\).
Write \(g(c)=1/(c-t)\) and \(h(c;m)=(1+tm)/c\). The map \(m\mapsto\int\max\{g(c),h(c;m)\}\,dF(c)\) preserves \([g(b),g(a)]\), since \(h\) decreases in cost, increases in the mean, and \(h(c;g(c))=g(c)\). Its Lipschitz constant in the mean is at most \(t/a<1\), so its fixed point \(\eta^*\) exists and is unique. Write \(\ell=\min\operatorname{supp}F\) and \(u=\max\operatorname{supp}F\). On \([g(u),g(\ell)]\), the map sends its lower endpoint strictly up, since \(g(c)>g(u)\) for \(c<u\). It sends its upper endpoint strictly down, since both branches are below \(g(\ell)\) for \(c>\ell\). Absolute continuity gives positive mass to both sets. Hence \(c^*\) is strictly between those extremes and
\[
 g(c)-h(c;\eta^*)=\frac{t\eta^*(c^*-c)}{c(c-t)} .
\]

\emph{Finite task comparisons.}
Write \(w(e)=e+te^2/2\). In an ingroup task put \(x=1/c\) and \(y=1/d\), where the two costs are \(c,d\). The equilibrium efforts solve \(e=x(1+tf)\), \(f=y(1+te)\). Thus \(e=x(1+ty)/(1-t^2xy)\), with the analogous formula for \(f\), and
\begin{align*}
 e-g(c)&=\frac{tx(y-x)}{(1-tx)(1-t^2xy)},\\
 g(c)+g(d)-e-f
 &=\frac{t(x-y)^2}{(1-tx)(1-ty)(1-t^2xy)} .
\end{align*}
Also \(ef/[g(c)g(d)]=(1-t^2x^2)(1-t^2y^2)/(1-t^2xy)^2\), where the denominator minus the numerator is \(t^2(x-y)^2\). Consequently, for some constant \(C\) independent of the pair,
\begin{equation}
 t(x-y)^2\le w(g(c))+w(g(d))-(e+f+tef)\le C(x-y)^2 .
 \label{eq:jten-pair-loss}
\end{equation}
Indeed, the middle term is the displayed effort-sum difference plus \(t[g(c)-g(d)]^2/2+t[g(c)g(d)-ef]\). All denominators are bounded away from zero. The individual absolute effort errors from \(g(c),g(d)\) sum to at most \(C|x-y|\), increasing \(C\) if needed. For a cross task the identity
\begin{equation}
 e+f+tef=w(e)+w(f)-t(e-f)^2/2
 \label{eq:jten-cross-loss}
\end{equation}
measures the score loss from unequal efforts.

\emph{A bound for every matching.}
Let \(F_{G,n}\) be group \(G\)'s empirical law and \(M\) the larger of the two actual group means. The ingroup effort-sum comparison and the outgroup best reply give
\[
 M\le\max_G\int\max\{g(c),h(c;M)\}\,dF_{G,n}(c).
\]
The right side is \(t/a\)-Lipschitz in \(M\). At \(\eta^*\), its distance from \(\eta^*\) is at most
\[
 r_n=\max_G\left|\int e^*(c)\,dF_{G,n}(c)-\eta^*\right|\longrightarrow0,
\]
where convergence follows from continuity of \(e^*\). It follows that every matching satisfies \(M\le\overline\eta_n=\eta^*+r_n/(1-t/a)\). Write \(e_n^+(c)=\max\{g(c),h(c;\overline\eta_n)\}\), which converges uniformly to \(e^*\). Applying \eqref{eq:jten-pair-loss} to ingroup tasks and \eqref{eq:jten-cross-loss} to cross tasks bounds \(S(\phi)/\kappa\) above by \(\sum_i w(e_n^+(c_i))\). After division by \(n\), the upper bound on normalized success per task converges to \(2\int w(e^*(c))\,dF(c)\), since each group contributes \(n\) workers.

\emph{An assignment attaining the bound.}
Choose \(k\) nearest to \(n[1-F(c^*)]\) among \(0\le k\le n\) with \(n-k\) even. This changes the target by at most one, including when \(n\) is odd. In each group pair consecutive ranks among the lowest \(n-k\) costs and send the highest \(k\) across groups in corresponding ranks. Order equal costs arbitrarily. In either group, \(|k-\#\{i\in G:c_i>c^*\}|=o(n)\), since \(F\) has no atom at \(c^*\). This is precisely the number whose assignment differs from selecting costs strictly above \(c^*\) for outgroup work.

Adjacent ingroup pairs have total reciprocal-cost span at most \(1/a-1/b\) in each group. The pair estimates therefore give bounded total absolute effort error relative to \(g\), and bounded total ingroup score loss relative to \(\sum_{i\ {\rm ingroup}}w(g(c_i))\). If outgroup workers best-reply to the trial mean \(\eta^*\), each group's mean differs from \(\int e^*(c)\,dF_{G,n}(c)\) by \(o(1)\), since only \(o(n)\) workers use the other branch and all efforts are bounded. The coefficient on the other group's mean in either actual mean equation is \(tn^{-1}\sum_{i\ {\rm outgroup\ in}\ G}1/c_i\le t/a\). Consequently the largest absolute error of the two actual means from \(\eta^*\) is at most the largest error of the trial means divided by \(1-t/a\). Thus both actual means converge to \(\eta^*\).

To bound cross-task losses, write \(c_{G,(j)}\) for the \(j\)-th cost rank of \(G\). Integrating the number of rank pairs separated by \(s\) gives
\[
 \frac1n\sum_{j=1}^n|c_{G_1,(j)}-c_{G_2,(j)}|
    =\int_a^b|F_{G_1,n}(s)-F_{G_2,n}(s)|\,ds\ \longrightarrow 0 .
\]
The convergence follows by bounded convergence at the continuity points of \(F\). Compactness also gives convergence of the mean squared rank differences to zero, and any upper rank tail contributes no more than the entire sum. Since \(h\) is Lipschitz on the relevant cost and effort intervals and the two means converge to \(\eta^*\), \(\sum_{\rm cross}(e-f)^2=o(n)\). The absolute individual effort errors from \(e^*\) sum to \(o(n)\) by the preceding comparisons. Equation \eqref{eq:jten-cross-loss} and the ingroup estimates now show that the assignment attains the limiting upper bound.

\emph{Efforts and assignments in all optimal and near-optimal sequences.}
Consider any sequence with \((S_n^*-S(\phi_n))/(\kappa n)\to0\). The upper bound just proved and its attainment imply that \(\sum_i w(e_n^+(c_i))-S(\phi_n)/\kappa=o(n)\). Put \(s_i=g(c_i)\) ingroup and \(s_i=h(c_i;\overline\eta_n)\) outgroup. For brevity write \(e_i=e_i(\phi_n)\). The deficit is exactly
\begin{align*}
 &\sum_i[w(e_n^+(c_i))-w(s_i)]\\
 &\quad+\sum_{\{i,j\}\ {\rm ingroup}}
 [w(g(c_i))+w(g(c_j))-(e_i+e_j+te_ie_j)]\\
 &\quad+\sum_{i\ {\rm outgroup}}[w(s_i)-w(e_i)]
 +\tfrac{t}{2}\sum_{\{i,j\}\ {\rm cross}}(e_i-e_j)^2 .
\end{align*}
Each term is nonnegative and thus its sum is \(o(n)\). Since \(w'\ge1\), the first sum bounds the total difference between \(e_n^+\) and \(s_i\), and the third bounds the total outgroup difference between \(s_i\) and actual effort. By \eqref{eq:jten-pair-loss}, the sum of squared reciprocal-cost gaps in ingroup tasks is \(o(n)\). Cauchy--Schwarz makes the sum of their absolute gaps \(o(n)\), so the individual ingroup effort errors from \(g\) also sum to \(o(n)\). The errors are absolute, so the bounds hold separately for each group. In particular, for each \(G\), \(n^{-1}\sum_{i\in G}|e_i-e^*(c_i)|\to0\). Here uniform convergence of \(e_n^+\) to \(e^*\) replaces the empirical effort bound by its limit. Averaging gives convergence of \(\eta_G(\phi_n)\).

Outside any fixed positive margin around \(c^*\), the branch defining the strictly larger effort in \(e^*(c)\) has an advantage bounded away from zero wherever the set is nonempty. This comparison persists at \(\overline\eta_n\). A worker assigned the other branch there incurs a uniformly positive loss in the first sum. Hence there are \(o(n)\) such workers in each group. The limiting mass of a closed margin around \(c^*\) tends to zero with its width because \(F\) has no cutoff atom. Consequently the total proportion of assignment errors also vanishes in either group, with any convention for workers exactly at the cutoff. The outgroup fraction in each group tends to \(1-F(c^*)\). Each cross task uses one worker from each group, so this is also the limiting fraction of cross tasks.

As there are \(n\) tasks and \(p(u,v)=\kappa(u+v+tuv)\), the upper bound and its attainment yield
\[
 \lim_{n\to\infty}\frac{S_n^*}{n}
 =2\kappa\int_a^b w(e^*(c))\,dF(c)
 =\int_a^b p(e^*(c),e^*(c))\,dF(c).
\]
The same limit holds for the expected successes per task along any of the near-optimal sequences just considered.\hfill\(\square\)

\subsection{Full investment under rank matching}\label{proof:coordination-grid}
\noindent\textit{Proof of Proposition~\ref{prop:coordination-grid}.} Recall that \(x_k=k(1+\varepsilon)/n\), and the two players of each rank are paired across groups. Once all ranks above \(k\) invest, each member of pair \(k\) assigns probability at most \(k/n\)
to her partner not investing. Her gain from investing is
at least \(x_k-k/n=k\varepsilon/n>0\). Starting at \(k=n\) and working
down the ranks thus forces every player to invest in any OHE, including
mixed profiles. Full investment satisfies every player's best response. $\blacksquare$

\subsection{Multiple Bayesian equilibria on the investment grid}\label{proof:coordination-bayesian}
\noindent\textit{Proof of Proposition~\ref{prop:coordination-bayesian}.} When everyone invests, investment pays in every possible pair because
all pair means are positive. If an ingroup match has positive
probability, another BNE prescribes \(NI\) at every ingroup signal and
\(I\) after \(\mathrm{Out}\). Both partners observe the same match
status, and every ingroup pair has mean at most \((x_n+x_{n-1})/2<1\).

Suppose all matches are outgroup, and let \(p\) be the probability that
the two top players meet. If \(1-p\ge1/(2n)\), consider the profile in which no one invests.
Each top player's gain from investing is at most
\[
 x_n-1-\frac{1-p}{2}(x_n-x_{n-1})
 \le \varepsilon-\frac{1+\varepsilon}{4n^2}<0,
\]
and every other player's pair mean is below one. If \(1-p\le1/(2n)\),
let only the top players invest. Each meets an investor with probability
\(p>1/2\), and her pair mean always exceeds \(1/2\). For anyone else,
when the top players meet, the gain from investing is at most
\(x_{n-1}-1<-1/(2n-1)\). When the top players are apart, the gain is at
most \((x_n+x_{n-1})/2<1\). Hence this player's expected gain is below
\(1-p-p/(2n-1)\le0\). $\blacksquare$

\subsection{The investment cutoff at a fixed matching}\label{proof:coordination-cutoff}
\noindent\textit{Proof of Proposition~\ref{prop:coordination-cutoff}.} When neither member of the lowest \(k\) pairs invests and everyone else invests, the probability each player assigns to her partner choosing \(NI\) is \(k/n\). The gain from investing for either member of pair \(r\) is therefore \(m_r-k/n\). By the ordering of the pair means, this profile is an OHE precisely when the inequalities in the proposition hold, omitting the relevant endpoint comparison. Two distinct values of \(k\) satisfying them give distinct equilibria. To
cover all profiles, start with no one investing and iterate best
responses, choosing \(NI\) at indifference. Every player who switches
strictly prefers investment even if only the players already forced to
invest do so. By induction these choices give a lower bound on every
player's investment probability in every OHE. Starting from full
investment and choosing \(I\) at indifference gives the corresponding
upper bound. Both iterations treat the members of each pair alike and
assign investment to the pairs with the highest means. Each reaches a
cutoff OHE in finitely many steps, since investment increases in the
first iteration and decreases in the second. If exactly one \(k\)
satisfies the inequalities, the bounds coincide for every player. $\blacksquare$

\subsection{Multiple Bayesian equilibria for general interior lists}\label{proof:coordination-bayesian-general}
\noindent\textit{Proof of Proposition~\ref{prop:coordination-bayesian-general}.} Let \(\delta=\min\{x_2,1-x_{n-1}\}\) and take
\(\bar\varepsilon=\sqrt{1+\delta^2}-1\). Every ingroup pair then has
mean strictly between zero and one. If such a pair meets with positive
probability, its two identity signals can support either coordinated
action independently of play at all other signals. At all remaining signals, start with \(NI\) and iterate best responses,
retaining \(NI\) at indifference. This reaches a pure equilibrium in finitely many steps,
since investment incentives increase with other players' investment.
The two choices for the ingroup pair give distinct BNE.

For a prior supported on full-outgroup matchings, write \(p\) for
the probability that the top players meet. If no one invests, only
a top player can gain from investing. Her gain is at most
\(\varepsilon_H-(1-p)(\delta+\varepsilon_H)/2\). If this bound is
nonpositive, no player gains by deviating, so this is a BNE. Otherwise
\(1-p<2\varepsilon_H/(\delta+\varepsilon_H)<1/2\), since
\(\varepsilon_H<\delta^2/2\) and \(\delta\le1/2\).
Let only the top players invest. Each gains more than one when they
meet and more than minus one otherwise, so both prefer investment.
Any other player meets an investor with probability at most \(1-p\),
and only then can her partner's characteristic exceed \(1-\delta\).
Her investment gain is therefore at most
\[
  -\delta+(1-p)\left(1+\frac{\delta+\varepsilon_H}{2}\right)<0,
\]
where the inequality follows from
\(\varepsilon_H(2+\varepsilon_H)<\delta^2\).
This gives a BNE in which at most the top rank invests. Exchanging
\(I\) and \(NI\) and replacing \(x_i\) by \(1-x_i\) reverses the
investment gain and interchanges the endpoint excesses. The same
argument gives a BNE in which every player above the bottom rank invests.
Every interior rank acts differently in the two equilibria. $\blacksquare$

\subsection{Finite populations with a prescribed limiting investment cutoff}\label{proof:coordination-finite}
\noindent\textit{Proof of Corollary~\ref{cor:coordination-finite}.} Write \(k=\lfloor nc\rfloor\). Take the \(j/n\) quantiles of \(F\)
for \(j=1,\ldots,n-1\), omit the \(k/n\) quantile, and reserve the
first and last ranks for the two endpoints. For \(j<k\), the retained
quantile is at most \(j/n\) and occupies rank \(j+1\). For \(j>k\),
it is at least \(j/n\) and occupies rank \(j\). Move the lower interior
characteristics slightly down and the upper ones slightly up,
preserving their order and keeping them in \((0,1)\). Every lower
interior rank \(r\) now lies strictly below \((r-1)/n\), and every
upper interior rank \(r\) strictly above \(r/n\). If no one in the lowest \(r-1\) pairs invests, neither member of pair \(r\)
finds it worthwhile to invest. If everyone in the highest \(n-r\) pairs invests,
both members of pair \(r\) strictly prefer to invest. The two arguments
meet at \(k\).

Choose positive endpoint excesses below the bound in Appendix~\ref{proof:coordination-bayesian-general} and smaller
than \(1/n\). All the inequalities are strict and the bound depends
continuously on the interior list, giving the claimed neighborhoods.
Before perturbation, completing the quantile list with \(0\) and \(1\)
gives an empirical distribution function within \(1/n\) of \(F\)
uniformly. Taking all movements smaller than \(1/n\) preserves uniform
convergence by continuity of \(F\). $\blacksquare$

\section{Learning foundations}\label{app:learning}
\subsection{Cohorts, encounters, and action frequencies}\label{appendix:learning-foundation}

Fix the finite environment of Section~\ref{subsec:learning-foundation_ABEE}. The set of roles $I$ is nonempty and has even cardinality; $A$ is finite, common and nonempty; $x$ and the complete matching
$\phi$ are fixed; and the two labelled groups are disjoint and cover $I$.
An empty labelled group is allowed. The pair payoffs are real, and their
mixed extension is $U_i$. The physical consequences of an action profile may be correlated. The perturbations below enter subjects' payoffs; they leave the unperturbed game unchanged.

For reference, consistency at an aggregate profile $\sigma$ means
\begin{equation}
\label{eq:learning-consistency}
\beta_i=
\begin{cases}
\sigma_{\phi(i)}, & \phi(i)\in G(i),\\[1mm]
\bar\sigma_{G(\phi(i))}, & \phi(i)\notin G(i).
\end{cases}
\end{equation}
Every average used here is over a nonempty group containing the actual
partner. The average includes all its roles, including roles matched
ingroup. At each learning date the same expression is applied to the
preceding period's profile.

The following distributional population construction supplies exact
shares and an individual encounter law without a continuum law of large
numbers. Equip the finite product cube $[0,1]^{I\times A}$ with product
Lebesgue probability measure. In each period create a fresh mass-one
cohort for every role, each a copy of this cube. Coordinates are arrays
$(z_i(a))_{i\in I,a\in A}$. Of these coordinates, a subject in role $i$ observes only her own vector $(z_i(a))_{a\in A}$. She knows her payoff rule and observes her partner's role, characteristics and group membership, as well as the action frequencies described below. The full coordinate
array is a hidden accounting label, not an observable identity.

Within each fixed role pair, match copies carrying the same full
coordinate array. Each subject participates in exactly one encounter.
Given the preceding period's action frequencies, each subject's choice
depends only on her own shock vector. Since the role coordinates are
independent, so are the actions in a representative encounter. More generally, each full replica
has the product action law used by the baseline mixed extension.
Applying the original action-contingent physical lottery in each replica
retains its conditional physical distribution and its permitted
correlations. A subject who chooses $a$ receives the action-specific utility perturbation $\varepsilon z_i(a)$ in addition to baseline utility, with $\varepsilon>0$ fixed throughout the process.

Population action shares are the measures of the sets of coordinates
inducing each action. Each role has mass one and each subject chooses one
action. Hence the action frequencies pooled across a group equal the
arithmetic average of its role distributions. An arbitrary initial profile supplies the first period's frequencies. Before acting, each subject observes the preceding period's frequencies separately for every other role in her own group and only the pooled frequencies for the opposite group, when nonempty. She has no access to other roles' private observations,
disaggregated outgroup histories, or a label revealing her partner's
private coordinates.
Fresh subjects act once; their successors receive no additional personal
history. Reusing the coordinate space across periods is bookkeeping and
does not identify subjects across generations or assert a temporally
independent random field along each fixed coordinate label.

For a fixed conjecture and positive $\varepsilon$, choose a maximizing
action at each private shock vector. A fixed ordering of the finite action
set may resolve ties. This defines a measurable action rule. The proof below shows that ties have measure zero, so their treatment does not
affect the induced aggregate action distribution. Thus the conditional
pure action rule is recoverable almost everywhere from the conjecture;
the definition need not introduce an additional strategic object.

\subsection{Stationary profiles and vanishing payoff perturbations}\label{proof:learning-foundation}

\noindent\textbf{Proposition~\ref{prop:learning-foundation} (restated).} \emph{For every $\varepsilon>0$, the stationary aggregate profiles of the
specified learning process coincide with the payoff-perturbed OHE profiles
at scale $\varepsilon$, and this set is nonempty. Every accumulation point
of such profiles as $\varepsilon$ tends to zero is an OHE of the
unperturbed environment.}

\noindent\textbf{Proof.}
Fix $\varepsilon>0$. The first step is to show that the learning process defines a continuous aggregate response to every preceding-period profile.

For any fixed conjecture $\beta_i$, indifference between two distinct
actions imposes a single affine equality on the two corresponding shock
coordinates. Such a set is a Lebesgue-null hyperplane in $[0,1]^A$.
Because there are finitely many action pairs and the shock distribution
has a joint density, the maximizing action is unique almost surely.
If $A$ has one element, uniqueness is immediate instead. The choice
probabilities in \eqref{eq:payoff-perturbed-ohe-choice} therefore sum to
one and define a uniquely determined aggregate response.

To verify continuity, take a convergent sequence of preceding-period
profiles. The associated conjectures converge because each is a projection
or a finite arithmetic average. For each pure action, expected baseline
payoff consequently converges. At any shock vector at which the limiting
problem has a unique maximizer, its strict payoff margin over all other
actions persists for sufficiently advanced terms of the sequence.
The chosen-action indicators therefore converge almost everywhere.
They are bounded by one, so dominated convergence proves continuity of
every aggregate action probability.

The next-period response is thus a continuous self-map of the nonempty,
compact and convex finite-dimensional set $(\Delta(A))^I$. Brouwer's
fixed-point theorem gives a fixed point. Using this profile as the initial action distribution reproduces the same
observed frequencies and action shares in every
subsequent fresh generation, so it is a stationary aggregate profile.

At any stationary profile, the preceding and current action frequencies
coincide. The conjectures at that profile therefore satisfy
\eqref{eq:learning-consistency}, and its current action shares satisfy
\eqref{eq:payoff-perturbed-ohe-choice}. Hence it is a payoff-perturbed OHE.
Conversely, those two conditions imply that initializing the learning
process at a payoff-perturbed OHE reproduces that profile. The exact
population construction above implements these distributions in every
generation. This proves the equality of the two profile sets, as well as
their nonemptiness.

For the limit statement, fix any payoff-perturbed OHE at scale
$\varepsilon$, and denote its aggregate profile and consistent conjectures
by $\sigma$ and $\beta$. For each subject, perturbed optimality implies
that no alternative pure action improves her baseline expected payoff,
evaluated at $\beta_i$, by more than $\varepsilon$:
the difference between any two perturbation bonuses is at most
$\varepsilon$. Averaging over the subject's shock vector and then over
any mixed deviation gives
\begin{equation}
\label{eq:learning-regret-bound}
U_i(\sigma_i,\beta_i;x_i,x_{\phi(i)})
\geq
U_i(\sigma_i',\beta_i;x_i,x_{\phi(i)})-\varepsilon
\quad
\text{for every }i\in I,\ \sigma_i'\in\Delta(A).
\end{equation}
This is a bound on the unperturbed \emph{subjective} decision problem
evaluated at $\beta_i$, not a bound against the actual partner's
distribution when that differs from the conjecture.

Now let $\varepsilon_n>0$ tend to zero, and select a payoff-perturbed OHE
$\sigma^n$ at each scale $\varepsilon_n$. The index $n$ labels a sequence
of perturbed models, not learning periods. By compactness, every such
sequence has a convergent subsequence. Let $\sigma$ be any accumulation
point and restrict attention to a subsequence converging to it. The
corresponding consistent conjectures $\beta^n$ converge to the conjecture
profile $\beta$ given by \eqref{eq:learning-consistency} at $\sigma$.
For each player and every fixed mixed deviation $\sigma_i'$, continuity
of the finite mixed-payoff expression permits passage to the limit in
\eqref{eq:learning-regret-bound}. Thus
\[
U_i(\sigma_i,\beta_i;x_i,x_{\phi(i)})
\geq U_i(\sigma_i',\beta_i;x_i,x_{\phi(i)})
\qquad\text{for every }i\in I,\ \sigma_i'\in\Delta(A).
\]
Together with consistency, these are the unperturbed OHE conditions.
This proves the accumulation-point claim. $\blacksquare$

\section{Organizational benchmarks}\label{app:benchmarks}
\subsection{The optimal matching under Nash equilibrium}
\label{proof:org-nash-benchmark}

Consider the technology, fixed groups and unit bonuses of Section~\ref{section:orgs}.
There are \(2n\) workers and \(n\) two-person tasks. All normalized costs
\(c_i=x_i/\kappa\) lie in \([a,b]\), with \(a>t+1/\bar e\). Write
\(g(c)=1/(c-t)\) for the effort of a worker paired with an equally costly
teammate under Nash equilibrium. Let \(S_n^{\mathrm N}\) and \(S_n^*\)
denote the maximum expected number of successful tasks under Nash equilibrium
and OHE, respectively. In each case the designer chooses a matching.

\paragraph{Result.} When each worker observes her teammate, pairing consecutive
workers in the pooled cost ranking attains \(S_n^{\mathrm N}\).
Under the common-limit assumptions of Proposition~\ref{prop:large-finite-cutoff},
the average absolute difference between effort under this pairing and \(g(c_i)\)
tends to zero, and
\[
 \lim_{n\to\infty}\frac{S_n^{\mathrm N}}{n}
 =\int_a^b p(g(c),g(c))\,dF(c).
\]
The limiting OHE gain per task is strictly positive and comes entirely from
costs above the cutoff:
\[
 \lim_{n\to\infty}\frac{S_n^*-S_n^{\mathrm N}}{n}
 =\int_{(c^*,b]}\!\bigl[p(e^*(c),e^*(c))-p(g(c),g(c))\bigr]\,dF(c)>0 .
\]

\paragraph{Proof.}
In every team the two workers best respond to one another. Strict concavity
makes each best reply pure, and the slope of a best reply in teammate effort
is at most \(t/a<1\), so every matching has a unique Nash equilibrium.
A task with costs \(c,d\) therefore has the same success probability
\(P(c,d)\) as an ingroup task with the same costs under OHE.
This function is symmetric and supermodular, as shown in
Appendix~\ref{proof:org-supermodularity}. In the pooled cost list,
consider the first two remaining workers, with costs \(a'\le b'\).
If they work separately, let their respective partners have costs
\(u,v\ge b'\). Supermodularity gives
\[
 P(a',b')+P(u,v)\ge P(a',u)+P(b',v).
\]
Pairing the first two workers and pairing their two partners thus weakly
raises expected successes. Fix the first pair and repeat on the remaining
list. The argument also applies when costs tie.

Write \(w(e)=e+te^2/2\). For consecutive pairing, the finite task bound
\eqref{eq:jten-pair-loss} gives
\[
 0\le \sum_i w(g(c_i))-S_n^{\mathrm N}/\kappa
 \le C\sum_{\{i,j\}\,\text{consecutive pair}}(1/c_i-1/c_j)^2
 \le C(1/a-1/b)^2 .
\]
The constant \(C\) is independent of \(n\); consecutive reciprocal-cost
intervals have disjoint interiors. The individual effort estimate in
Appendix~\ref{proof:large-finite-cutoff} similarly bounds the total absolute
effort error from \(g(c_i)\) by \(C(1/a-1/b)\). Dividing by \(2n\) proves
the effort claim. Dividing the success bound by \(n\) and using convergence
of each group's empirical law gives the limit for the Nash optimum, since
\(2\kappa w(e)=p(e,e)\). By Appendix~\ref{proof:large-finite-cutoff},
the OHE optimum per task tends to \(\int_a^b p(e^*(c),e^*(c))\,dF(c)\).
At costs at or below \(c^*\), \(e^*(c)=g(c)\). Above \(c^*\),
\(e^*(c)>g(c)\), and \(p\) strictly increases in each effort.
The latter region has positive \(F\)-mass, which gives the strict
inequality.\hfill\(\square\)

\subsection{Bayesian effort and expected task success}
\label{proof:org-bayesian-value}

Retain the finite organization just described. In the random-matching
game of Section~\ref{sec:benchmark-matching}, an ingroup signal identifies
the teammate and an outgroup signal reveals only \(\mathrm{Out}\). The designer
chooses a common prior on perfect matchings; the signal rule stays fixed.

\paragraph{Result.} For every prior the BNE is unique at signals of positive
probability. For a worker drawn uniformly and independently of the matching,
let \(C\) denote her cost, \(X\) her effort and \(Y\) her actual teammate's
effort. Then
\[
 \mathbb E[X]\le \mathbb E[g(C)],\qquad
 \frac{S_n^{\mathrm N}}{n}
 \le \sup_{\mu\in\Delta(\Phi)}\mathbb E_\mu[p(X,Y)]
 \le \mathbb E[p(g(C),g(C))].
\]
The first inequality holds under every prior; in the second comparison the
expectation with subscript \(\mu\) uses the equilibrium for that prior.
Under the assumptions of Proposition~\ref{prop:large-finite-cutoff},
the supremum of Bayesian expected successes per task converges to
\(\int_a^b p(g(c),g(c))\,dF(c)\). Thus the limiting OHE gain over this
benchmark equals the strictly positive gain in Appendix~\ref{proof:org-nash-benchmark}.

\paragraph{Proof.}
Under~\eqref{eq:org-interiority}, the best reply at each possible signal is
pure and interior. On the worker--signal pairs of positive probability,
posterior averages are nonexpansive, so the best-response map contracts
the effort box by a factor of at most \(t/\min_i c_i<1\). This gives uniqueness.

Let \(H=g(C)\). Conditional on any matching, a uniform worker has a uniform
teammate, so \(X\) and \(Y\) have the same marginal law. With \(\mathcal I\)
denoting the selected worker's identity and signal, the best reply is
\[
 X/H=1+t\bigl(\mathbb E[Y\mid\mathcal I]-X\bigr).
\]
For \(k=1,2\), multiply by \(X^k\) and take expectations. Since \(X\) is
\(\mathcal I\)-measurable and H\"older's inequality gives
\(\mathbb E[X^kY]\leq\mathbb E[X^{k+1}]\) by equality of the marginals,
\(\mathbb E[X^{k+1}/H]\leq\mathbb E[X^k]\).
A second application of H\"older gives
\[
 \mathbb E[X^k]\leq
 \bigl(\mathbb E[X^{k+1}/H]\bigr)^{k/(k+1)}
 \bigl(\mathbb E[H^k]\bigr)^{1/(k+1)},
 \qquad\text{hence}\qquad
 \mathbb E[X^k]\leq\mathbb E[H^k].
\]
Taking \(k=1\) bounds mean effort. Each task is counted twice when its
two workers are counted separately, so \(\mathbb E[p(X,Y)]\) is the expected
success per task. Combining \(\mathbb E[XY]\leq\mathbb E[X^2]\) with both
moment bounds gives
\[
 \mathbb E[p(X,Y)]\leq\kappa\mathbb E[2H+tH^2]=\mathbb E[p(H,H)] .
\]
For the lower bound, concentrate the prior on a Nash-optimal matching.
Every worker can then infer her teammate's identity, even if her signal is
\(\mathrm{Out}\), so the equilibrium attains \(S_n^{\mathrm N}/n\).
Under the common-limit assumptions the lower bound converges by
Appendix~\ref{proof:org-nash-benchmark}. The upper bound has the same limit
by weak convergence of the cost laws and continuity of \(p(g(c),g(c))\).
This establishes the Bayesian comparison.\hfill\(\square\)

\clearpage
\pdfbookmark[0]{References}{paper-references}
\bibliographystyle{plainnat}
\bibliography{refs}

\clearpage
\pdfbookmark[0]{Online Appendix}{online-appendix}
\section*{Online Appendix}
\setcounter{section}{0}
\renewcommand{\thesection}{\Roman{section}}
\renewcommand{\theHsection}{online.\Roman{section}}
\renewcommand{\theHsubsection}{\theHsection.\arabic{subsection}}
\renewcommand{\theHsubsubsection}{\theHsubsection.\arabic{subsubsection}}
\numberwithin{proposition}{section}
\numberwithin{definition}{section}
\renewcommand{\theHproposition}{online.prop.\Roman{section}.\arabic{proposition}}
\renewcommand{\theHdefinition}{online.def.\Roman{section}.\arabic{definition}}
\section{A limitation of the learning result}
\label{oa:learning-scope}

In the learning environment of Proposition~\ref{prop:learning-foundation}, successive cohorts play the same finite-action game at a fixed matching. Each subject observes her private payoff shocks. She uses the preceding cohort's action frequencies in her current partner's role if matched ingroup, and in her partner's entire group if matched outgroup. A stationary profile reproduces the action frequencies on which these conjectures are based. The proposition shows that any limit of stationary profiles as the shocks vanish is an OHE. The argument permits arbitrary real finite payoff matrices, unequal
groups, a labeled empty group, mixed ingroup/outgroup pairings, and a
single available action. It imposes neither private values nor a
full-outgroup matching. No positive action-share, uniqueness, dynamic
convergence, attraction, or selection conclusion follows.

The reverse limit claim fails in general for the fixed uniform-shock
specification. For example, let $A=\{0,1\}$ and let all baseline payoffs
be zero. Every aggregate action profile is then an unperturbed OHE.
For every positive perturbation scale, however, each subject chooses the
action with the larger of two independent uniform shocks. Each action
therefore has probability $1/2$, whatever the subject's conjecture about her partner's action.
The unique payoff-perturbed OHE is the profile assigning $1/2$ to each
action in every role, so the other OHE profiles are not its vanishing-noise
limits. The proposition deliberately asserts only the forward inclusion.

\section{Equilibrium comparisons in the organizational model}
\subsection{Random matching with bounded effort}
\label{oa:bounded-effort-bayesian}
Two fixed groups of workers are matched in pairs. Workers choose effort simultaneously, and a successful task pays each teammate a unit bonus. Let \(A=[0,\bar e]\), where \(\bar e>0\), and suppose
that \(x_i>0\) for every \(i\in I\) and \(t\in(0,1)\). The payoff is
\[
u_i(e_i,e_j;x_i,x_j)
=\kappa(e_i+e_j+t e_i e_j)-\frac{x_i}{2}e_i^2,
\qquad
0<\kappa\le (2\bar e+\bar e^2)^{-1}.
\]
Here \(\Delta(A)\) denotes the Borel probability measures on \(A\).
For \(\alpha,\gamma\in\Delta(A)\), define
\[
U_i(\alpha,\gamma;x_i,x_j)
=\int_A\int_A
u_i(e,z;x_i,x_j)\,\gamma(dz)\,\alpha(de).
\]
The Bayesian information structure and the OHE consistency and
optimality conditions are as in the finite-action comparison,
with these mixed payoffs. In particular, the fixed primitives are
common knowledge in the Bayesian benchmark. Write \(\Phi_{\mathrm{Out}}\) for the set of perfect matchings across groups. Players receive
only \(\mathrm{Out}\) before acting when the prior is supported on
\(\Phi_{\mathrm{Out}}\). Conditional on the realized matching,
players' action draws are independent. At the fixed matching under OHE, each worker observes her teammate's
identity and both workers' effort costs.

\begin{proposition}
\label{prop:ohe-bne-effort-equivalence}
Suppose \(|G_1|=|G_2|=n\ge1\), and let \(\mu\) be the uniform
distribution on \(\Phi_{\mathrm{Out}}\). For every
\(x\in(0,\infty)^I\), every \(\varphi\in\Phi_{\mathrm{Out}}\), and
every profile \(\sigma\in(\Delta(A))^I\), \(\sigma\) is an OHE
under the fixed matching \(\varphi\) if and only if the decision
functions \(s_j(\mathrm{Out})=\sigma_j\), \(j\in I\), form a BNE
under \(\mu\).
\end{proposition}

\textbf{Proof.}
Fix \(x\), \(\varphi\in\Phi_{\mathrm{Out}}\), and \(\sigma\), and
define \(s_j(\mathrm{Out})=\sigma_j\) for every \(j\in I\).
For each player \(i\), the payoff is continuous on \(A^2\) and
satisfies
\[
|u_i(e,z;x_i,x_j)|
\le \kappa(2\bar e+t\bar e^2)+\frac{x_i}{2}\bar e^2
\qquad\text{for all }e,z\in A.
\]
Thus, all mixed payoffs are finite. A finite mixture of Borel
probability measures on \(A\) is again a Borel probability measure,
and linearity of integration makes \(U_i\) affine in the partner's
mixed action.

Every player receives \(\mathrm{Out}\) with probability one.
There are \(n!\) feasible outgroup matchings and \((n-1)!\) that
assign any given outgroup partner to a given player. Therefore,
for every \(i\in I\) and every \(\sigma_i'\in\Delta(A)\),
\[
\begin{aligned}
\bar U_i(\sigma_i',s_{-i};\mathrm{Out})
&=\sum_{\psi\in\Phi_{\mathrm{Out}}}\mu(\psi)\,
U_i(\sigma_i',\sigma_{\psi(i)};x_i,x_{\psi(i)})\\
&=\frac1n\sum_{j\in I\setminus G(i)}
U_i(\sigma_i',\sigma_j;x_i,x_j)\\
&=U_i\!\left(
\sigma_i',\bar\sigma_{I\setminus G(i)};
x_i,x_{\varphi(i)}
\right).
\end{aligned}
\]
The last equality uses private values, which this payoff
satisfies, and linearity of integration with respect to the
finite mixture \(\bar\sigma_{I\setminus G(i)}\).

Under \(\varphi\), consistency uniquely determines
\(\beta_i=\bar\sigma_{I\setminus G(i)}\) for every \(i\).
Applying the identity to \(\sigma_i\) and every alternative
\(\sigma_i'\) shows that
\[
\bar U_i(\sigma_i,s_{-i};\mathrm{Out})
\ge \bar U_i(\sigma_i',s_{-i};\mathrm{Out})
\]
holds if and only if
\[
U_i(\sigma_i,\beta_i;x_i,x_{\varphi(i)})
\ge U_i(\sigma_i',\beta_i;x_i,x_{\varphi(i)}).
\]
These are precisely the Bayesian and OHE optimality conditions.
Together with the consistent conjectures just constructed, this
proves both directions.
\(\blacksquare\)

\subsection{When are cross-cost effects strict?}\label{oa:strict-cost-effects}

Fix a matching \(\varphi\), the groups, the effort cap \(\bar e\), and the technology parameters \(t,\kappa\) of Online Appendix~\ref{oa:bounded-effort-bayesian}.
Assume \(c_i=x_i/\kappa>t+1/\bar e\) for every worker, so each best reply is interior. Write \(w_{ij}\) for worker \(i\)'s weight on worker \(j\)'s effort: in an ingroup match, it is one for her teammate and zero otherwise; in an outgroup match, it is the reciprocal of the size of her teammate's group for every member of that group and zero otherwise. Thus \(c_i e_i=1+t\sum_j w_{ij}e_j\). Here the weights determine expectations, while task success depends on the efforts of actual teammates. The objective at equilibrium is \(S(\varphi)=\kappa\sum_i e_i+(\kappa t/2)\sum_i e_i e_{\varphi(i)}\).

If all tasks are ingroup, the mixed cost derivative of \(S(\varphi)\) is strictly positive within each team and zero across teams. If there is at least one outgroup task, it is strictly positive for every two distinct workers' costs.

Smoothness follows from
the finite linear system:
\[
e=\bigl(\operatorname{diag}(c)-tW\bigr)^{-1}\mathbf1,
\qquad W=(w_{ij}).
\]
The inverse exists because
\(\|t\operatorname{diag}(c)^{-1}W\|_\infty
=\max_i t/c_i<1\). Each effort is therefore smooth. Expanding the inverse as a geometric series expresses each effort as a sum of monomials in inverse costs with nonnegative coefficients. Thus each effort decreases weakly in every cost, and each finite partial sum satisfies the four-value inequality for supermodularity; the same inequality holds at the pointwise limit (Appendix~\ref{proof:org-supermodularity}). These properties and smoothness imply
\(\partial_j e_i\le0\) and \(\partial_{jk}e_i\ge0\) for \(j\ne k\).
For any two workers \(a,b\),
\[
\partial_{jk}(e_a e_b)
=e_b\partial_{jk}e_a+e_a\partial_{jk}e_b
+(\partial_j e_a)(\partial_k e_b)
+(\partial_k e_a)(\partial_j e_b)\ge0.
\]
For an ingroup pair, both efforts decrease strictly in both costs,
as the terms \(1/c_a\) and \(t/(c_ac_b)\) and their counterparts
show. More precisely, subtracting a selected finite monomial from
the expansion leaves a smooth nonincreasing function, so the strict
sign follows without termwise differentiation. The displayed product
rule consequently gives a strictly positive mixed cost derivative
within that team. When all tasks are ingroup, efforts depend only on
the costs of the actual teammates, so mixed derivatives across teams
are zero.

If an outgroup task \(\{a,b\}\) exists, with \(a\in G\) and \(b\in G'\),
the expansion for \(e_a\) contains
\[
\frac{t}{|G'|c_ac_j}\quad(j\in G'),
\qquad
\frac{t^2}{|G'||G|c_ac_bc_j}\quad(j\in G).
\]
The second term comes from the path \(a\to b\to j\). Subtracting
the relevant monomial again leaves a smooth nonincreasing function.
Thus both outgroup teammates' efforts have strictly negative first
derivatives in every cost. For every distinct \(j,k\), the last two
terms in the product rule are strictly positive and the other terms
are nonnegative. Since \(\kappa t e_a e_b\) enters the objective and
all remaining terms have nonnegative mixed cost derivatives, the
objective has strictly positive mixed derivatives between every pair
of distinct costs. \(\square\)

\section{Changing two tasks in a finite organization}\label{oa:local-rematches}

Keep the groups and bonuses fixed. A matching assigns each worker to one two-person task. At efforts \(e,f\), the task succeeds with probability \(p(e,f)=\kappa(e+f+tef)\); each worker receives one if it succeeds, and worker \(i\) pays effort cost \(x_i e^2/2\). Assume \(0<t<1\), \(\bar e>0\), \(0<\kappa\le(2\bar e+\bar e^2)^{-1}\), and normalized costs \(c_i=x_i/\kappa>t+1/\bar e\). Best replies are interior: a worker expecting effort \(z\) from her teammate chooses \((1+tz)/c_i\). Ingroup, \(z\) is her actual teammate's effort; outgroup, it is mean effort across the teammate's entire group, including its ingroup workers. The designer compares \(S(\phi)\) and \(S(\psi)\), the expected numbers of successful tasks in the respective equilibria. Only the four selected workers change teammates.

Write \(m_k=\eta_{G_k}(\phi)\) and \(m'_k=\eta_{G_k}(\psi)\) for the old and new group means, \(\delta_k=m'_k-m_k\), and \(N_k=|G_k|\).

\subsection{Opening two ingroup tasks when both original means are high}
\label{app:local-rematches}

Under \(\phi\), workers \(i_k,j_k\) form an ingroup team in \(G_k\), with \(c_{i_k}\le c_{j_k}\). Under \(\psi\), the two cheaper workers team up across groups, as do the two dearer workers. Suppose the cost intervals overlap: \(\max\{c_{i_1},c_{i_2}\}\le\min\{c_{j_1},c_{j_2}\}\). If the smaller original group mean weakly exceeds all four original efforts, the rematch raises expected successes strictly unless all four original efforts and both means coincide.

For outgroup tasks, hold expected teammate effort at the original mean of the teammate's group. Let $a_k$ be the resulting change in the
sum of the two selected workers' efforts in $G_k$. The efforts used here for
workers assigned ingroup under $\psi$ are their mutual best replies.
Put $A_k=\sum_{i\in\mathrm{Out}^{\psi}_{G_k}}1/c_i$. Every worker
in $G_k$ assigned outgroup under $\psi$ adds $t\delta_\ell/c_i$ to her effort when
expected teammate effort changes from $m_\ell$ to $m'_\ell$, where $\ell\ne k$.
All other ingroup pairs retain their efforts. Consequently
\[
 N_1\delta_1=a_1+tA_1\delta_2,\qquad
 N_2\delta_2=a_2+tA_2\delta_1.
\]
Since $tA_k<N_k$, the determinant is positive and the inverse of this
two-equation system has nonnegative entries. In particular,
$a_1,a_2\ge0$ imply $\delta_1,\delta_2\ge0$.

If the smaller original group mean weakly exceeds all four original efforts,
best-replying to this common mean weakly raises each selected effort.
At these trial efforts, cost overlap puts the cheaper worker of each original
pair among the two highest-effort workers, allowing ties. Pairing the two
cheaper workers together and the two dearer workers together then weakly
increases the sum of products by the rearrangement inequality. Replacing the common trial mean by the original mean of each worker's
opposite group also weakly increases each effort. The mean equations above now imply that both
equilibrium means weakly rise. A strict gain occurs if any of the four trial
efforts rises strictly or if either original mean exceeds the common trial
mean. Equality therefore requires all four old efforts and both old means
to coincide. In that case the old efforts still solve every best-reply
equation under $\psi$, so the two matchings yield the same success.

\subsection{Higher task success with lower total effort}\label{oa:opening-example}
Let $t=1/2$, $\bar e=3$ and $\kappa=1/15$. In each group an ingroup pair
has costs $(1,2)$. A second ingroup pair has common cost $177/130$ in
$G_1$ and $333/218$ in $G_2$. One further worker of cost $2$ in each group
works with her counterpart in the other group. Rematch the two $(1,2)$
pairs outgroup, pairing equal costs. The common-cost pairs keep their
teammates. Substitution in the pair and mean equations gives
\[
 (m_1,m_2)=(15/14,1),\quad
 (\delta_1,\delta_2)=(-3/448,1/448),\quad
 \frac{S(\psi)-S(\phi)}{\kappa}=\frac{611543}{3211264}.
\]
Total effort falls by $5/224$. In this example the original $(1,2)$
pair straddles the cost of the other ingroup pair in each group. The
comparison illustrates how success can rise when total effort falls.

\subsection{Closing two outgroup tasks when both means weakly rise}
\label{proof:closing-separated}

Starting with two outgroup teams, form one ingroup team in each group. Suppose the two selected workers in one group have strictly lower costs than either selected worker in the other group. If neither group's equilibrium mean falls, this rematch strictly raises the expected number of successful tasks. The calculation allows both means to stay fixed.

For this proof only, rescale efforts and group means by multiplying by
$t$, replace each cost $c$ by $c/t$, and multiply success by $t/\kappa$.
Suppressing the scale factors in the notation, the pair best reply is $(1+f)/c$, each cost exceeds one, and
each task contributes $e+f+ef$. Use $B$ for the group with the two
cheaper selected workers and $D$ for the other group. In these scaled coordinates, $\delta_B,\delta_D$ denote the changes in
the group means, and $N_B,N_D$ the group sizes. Let $E,F$ be the
new selected effort totals in $B,D$; let $P$ be the sum of the products on
the two new ingroup tasks. The old selected outgroup efforts were proportional
to inverse cost in each group. Let $\omega$ be the sum of the two old
cross products of inverse costs divided by the product of the two groups'
selected inverse-cost sums. Thus $0<\omega<1$ and, at selected totals
$E,F$ allocated in the old proportions and pairing, the cross products
sum to $\omega EF$. Put $M=P-\omega EF$ and $d=E-F>0$.

Two estimates will be useful:
\[
 M>0,\qquad 4M\ge d(d-1).
\]
To see the first, allocate each new pair total to its two workers in
proportion to inverse cost. For costs $c\le c'$ the resulting efforts are
$k/c,k/c'$, where $k=[1+2/(c+c')]/[1-1/(cc')]$ decreases in each
cost. Thus the two auxiliary efforts in $B$ strictly exceed the two
in $D$. Pairing them within groups gives a strictly greater sum of products
than either cross pairing. At the same pair total, switching to actual
ingroup efforts weakly increases each product again, since the effort ratio
$(c'+1)/(c+1)$ lies between $1$ and the auxiliary ratio $c'/c$.

For the second estimate, let the actual ingroup efforts be $x\ge x'$ in
$B$ and $y\ge y'$ in $D$, with differences $r=x-x'$, $s=y-y'$.
The effort gap for costs $c\le c'$ is $(c'-c)/(cc'-1)<1$.
Cost separation gives $(1+x)/x'\le (1+y')/y$, hence
$d\ge r(1+F)+s(1+E)$.
Write $\theta=(c'-c)/(c+c')$ for a pair's cost contrast. If its
total and gap are $E_0,r_0$, direct substitution gives
$\theta^2E_0\le r_0$: for $c<c'$ the ratio is
$(c'-c)(c+c'+2)/(c+c')^2<1$, and at equality both sides vanish.
Since $2\omega-1=\pm\theta_B\theta_D$ according to the old pairing,
\[
 2(2\omega-1)EF\le 2\sqrt{rsEF}\le rF+sE\le d-r-s .
\]
Using $4P=E^2+F^2-r^2-s^2$, this yields
$4M\ge d(d-1)+r(1-r)+s(1-s)\ge d(d-1)$.

Now suppose both group means weakly rise. Write $a,h$ for the changes in
the selected totals in $B,D$, so the original totals were $E-a,F-h$.
Let $A_B,A_D$ be the sums of the scaled inverse costs of the remaining
outgroup workers. The mean balances give
$a=N_B\delta_B-A_B\delta_D$ and $h=N_D\delta_D-A_D\delta_B$.
Since each group has at least two more workers than there are remaining
outgroup tasks, $A_D<N_B$ and $A_B<N_D$. Thus $a+h\ge0$.
Every remaining task weakly gains. If $a,h\ge0$, both the linear and
product terms on the selected tasks gain in total, with the change in the
product sum at least $M>0$. If $a=-L<0$, then $h\ge L$, and the change
in the product sum is $M+\omega(Eh-FL+Lh)>0$.

Finally, suppose $h=-L<0$. Then $0<L<a<E$ by the mean balances,
and the change in total organizational effort is
$N_B\delta_B+N_D\delta_D=a+(N_D+A_B)\delta_D\ge a$.
The products on remaining tasks weakly rise. Hence the gain in scaled
success is strictly greater than
\[
 a+P-\omega(E-a)(F+a)=M+a-\omega a(d-a).
\]
For $a\ge0$, $\omega a(d-a)-a\le (d-1)_+^2/4$, by maximizing
$a(d-1-a)$ over $0\le a\le d$ and observing that the left side
is negative for $a>d$. The two estimates above imply
$M>(d-1)_+^2/4$: use $M>0$ if $d\le1$, and
$d(d-1)>(d-1)^2$ otherwise. This proves strict improvement in all
cases. If both group means are fixed, the mean balances give $a=h=0$,
so the entire gain is $M$ in scaled units.

\subsection{Closing two outgroup tasks when one mean falls}\label{oa:closing-example}
Use the original effort and cost units. Replace two outgroup tasks with two ingroup tasks, leaving all other teammates fixed. Let $D$ be the
change in success on the selected tasks, evaluated at the old means
for the original outgroup tasks and at mutual best replies for the
new ingroup tasks. On the remaining outgroup tasks, adjusting the expected effort of a teammate from the second group and then that of a teammate from the first gives
\[
\begin{aligned}
 \frac{S(\psi)-S(\phi)}{\kappa t}
 ={}&\frac{D}{\kappa t}
 +\delta_2\sum_{i\in\mathrm{Out}^{\psi}_{G_1}}
 \frac{1}{c_i}\left(1+t\frac{1+tm_1}{c_{\psi(i)}}\right)\\
 &+\delta_1\sum_{j\in\mathrm{Out}^{\psi}_{G_2}}
 \frac{1+t e_{\psi(j)}(\psi)}{c_j}.
\end{aligned}
\]
Each term follows from the linear best reply and the affinity of task success
in either teammate's effort when the other's effort is held fixed. For a fall in $m_1$ and a weak rise in $m_2$, the switch is
strictly profitable exactly when the direct gain and the contribution from
the rise in $m_2$ together exceed the loss on the remaining outgroup tasks
from the fall in $m_1$. If no outgroup tasks remain, both sums vanish,
so $D$ is the entire gain. For instance, take $t=1/2$, $\bar e=3$,
$\kappa=1/15$, and two workers of cost $1$ in $G_1$ and two of cost $2$
in $G_2$. Forming two ingroup pairs from two outgroup pairs changes the
means from $(10/7,6/7)$ to $(2,2/3)$. Direct substitution gives
$(S(\psi)-S(\phi))/(\kappa t)=1552/441>0$.

\section{The optimal matching in the six-worker example}\label{proof:six-optimum}

In Example~\ref{ex:org-effort-spillover}, each group contains two workers of normalized cost \(c_L\) and one of cost \(c_H>c_L\). Under the interior assumptions of Online Appendix~\ref{oa:local-rematches}, a worker best responds to the effort of her actual ingroup teammate or to the mean effort of her outgroup teammate's entire group. The designer chooses three pairs to maximize the expected number of successful tasks. Pair the two low-cost workers in each group and assign the two high-cost workers to the same outgroup task. Write \(e_L=1/(c_L-t)\), \(e_H=(1+t\eta)/c_H\) and \(\eta=(2e_L+e_H)/3\) for their efforts and the common group mean. This matching is the unique optimum.

Use \(g(c)=1/(c-t)\), \(h(c;m)=(1+tm)/c\) and
\(w(e)=e+te^2/2\). In the illustrated matching,
\(e_L>\eta>e_H>g(c_H)\): the last inequality follows from \(e_H-g(c_H)=2t[e_L-g(c_H)]/(3c_H-t)>0\), and \(e_L-e_H=3(c_H-c_L)e_L/(3c_H-t)>0\).
Consequently \(g(c_L)>h(c_L;\eta)\) and
\(h(c_H;\eta)>g(c_H)\). Averaging the larger branch at each cost
therefore returns \(\eta=(2e_L+e_H)/3\). Under any other matching
the larger group mean \(M\) is at most the same average of the maxima
evaluated at \(M\): the effort sum of an ingroup pair of costs \(c,d\) is at most \(g(c)+g(d)\), by the finite task comparison in Appendix~\ref{proof:large-finite-cutoff}, while an outgroup worker of cost \(c\) exerts at most \(h(c;M)\). That average
is increasing and has Lipschitz constant at most \(t/c_L<1\), so
\(M\leq\eta\). Equations~\eqref{eq:jten-pair-loss}
and~\eqref{eq:jten-cross-loss} now bound total success divided by
\(\kappa\) by \(4w(e_L)+2w(e_H)\). The illustrated matching attains
the bound, since partners exert equal efforts in every task.
Sending a low-cost worker outgroup or keeping a high-cost worker
ingroup uses the strictly smaller branch for that worker in the bound.
It gives strictly less total success. The indicated statuses determine
all three pairs.\hfill\(\square\)

The same example also illustrates the Bayesian value comparison in Appendix~\ref{proof:org-bayesian-value}. A prior assigning probability one to the matching just described pairs equal-cost workers on every task. Each worker can infer her teammate's identity and exerts \(g(c)\). This attains the finite Bayesian upper bound on expected successes. Under OHE the two low-cost tasks have the same success probability, while the high-cost task has a strictly higher one because \(e_H>g(c_H)\).

\section{Pairwise stability at current efforts}\label{oa:stability}

Workers belong to two fixed groups and work in pairs under the interior effort assumptions of Online Appendix~\ref{oa:local-rematches}. Fix a matching and its OHE effort profile. A worker evaluating any prospective teammate uses that worker's current effort if they belong to the same group and the current mean effort of the entire opposite group otherwise. She holds her own effort fixed and applies the same rule to her incumbent teammate. Two workers on different tasks block the matching if both strictly prefer each other to their current teammates. No other worker's consent is required. A matching is stable if it has no such pair.

\subsection{Symmetric stable assignments and the limiting cutoff}\label{proof:decentralized-cutoff}

Consider two identical lists of \(n\) normalized worker costs, one per group. A symmetric assignment has the same cost pairs on ingroup tasks and the same list of outgroup costs in both groups, counting multiplicities. A symmetric stable matching exists for every such pair of lists. For the limiting claim, all costs lie in the same fixed \([a,b]\), with \(a>t+1/\bar e\), and the common empirical cost law converges weakly to an absolutely continuous \(F\). For every sequence of symmetric stable assignments, the cutoff of Proposition~\ref{prop:large-finite-cutoff} describes all but a vanishing fraction of workers in each group. Stability leaves the pairing of outgroup workers undetermined.

Write \(g(c)=1/(c-t)\), \(h(c;m)=(1+tm)/c\) and \(e^*(c)=\max\{g(c),h(c;\eta^*)\}\), where \(\eta^*=\int e^*(c)\,dF(c)\) and \(c^*=t+1/\eta^*\). For a symmetric assignment
the two group means coincide by uniqueness of the OHE; denote their
value by \(m\). At fixed own effort \(e_i\), changing expected teammate effort by \(d\) changes worker \(i\)'s subjective expected payoff by
\(\kappa(1+t e_i)d\). Thus a pair can block only if each worker expects strictly more effort
from the new teammate than from the current one.

\paragraph{Existence on each identical pair of finite lists.}
Start with all workers outgroup. While at least two remaining workers
in each group have \(g(c)>m\), pair the two cheapest of them ingroup
in each group and update the common mean. This constructs a matching;
it need not describe a sequence of mutually preferred moves.
Suppose the new pair has costs \(x\le y\) and efforts \(U\ge V\), with
\(U=h(x;V)\) and \(V=h(y;U)\). Then
\(g(x)\ge U\ge V\ge g(y)>m\) and
\(V-g(y)=t(U-V)/(y-t)\). In particular, every effort in an earlier
pair is at least every effort in the new pair.

To see that the updated mean stays below \(V\), let \(A(z)\) be the
mean implied by the old assignments when outgroup workers respond to
the trial mean \(z\). It is affine with fixed point \(m\) and slope
\(\beta=n^{-1}\sum_{i\ {\rm outgroup}}t/c_i<1\), using one group's
list. At \(z=V\), one new ingroup worker still exerts \(h(x;V)=U\),
and the other raises effort by \(t(U-V)/y\). Hence
\[
 A_{\rm new}(V)-V=(1-\beta)(m-V)+\frac{t(U-V)}{ny}<0 .
\]
Indeed \(m<g(y)\), \(n(1-\beta)\ge 1-t/y\), and
\((1-t/y)(V-g(y))=t(U-V)/y\). The new affine map also has slope
less than one, so its fixed point lies below \(V\).

At termination every ingroup worker and her teammate exert at least
the final mean. At most one outgroup worker per group exerts more than
that mean, since \(h(c;m)>m\) exactly when \(g(c)>m\). If such a
worker has cost \(z\) and there is an ingroup pair, then
\(h(z;m)<g(z)\le g(y)\) for the most costly ingroup member \(y\).
No ingroup worker wants to replace her teammate by this worker, by an
outgroup worker exerting at most the mean in her own group, or by a
prospect across groups. Two workers on distinct ingroup tasks in the
same group cannot both gain because the cheaper pair's two efforts
are at least the dearer pair's. Two outgroup workers in the same
group cannot both gain since at most one exceeds the mean; across
groups an incumbent outgroup worker is indifferent. These comparisons
also cover termination with all workers outgroup. This proves stability.

\paragraph{Necessity of the limiting cutoff.}
Consider a symmetric stable matching, not necessarily one constructed
above. Every ingroup worker's teammate must exert at least \(m\):
otherwise that worker and her counterpart on the corresponding ingroup
task in the opposite group would strictly prefer one another.
Thus every ingroup effort is at least \(m\). As above, at most one
outgroup worker per group can exert more than \(m\).
Within a group, the intervals between the two efforts on distinct
ingroup tasks have disjoint interiors. Otherwise the higher-effort
member of each pair would strictly prefer the other to her own
teammate. The sum of the lengths of these intervals is therefore at
most \(\bar e\). On each ingroup task \(\{i,j\}\),
\[
 e_i-g(c_i)=\frac{t}{c_i-t}(e_j-e_i).
\]
Hence the sum of \(\lvert e_i-g(c_i)\rvert\) over the ingroup workers of
either group is at most \(2t\bar e/(a-t)\), independent of group size.

Let \(E(c;m)=\max\{g(c),h(c;m)\}\). For an ingroup worker,
\(h(c_i;m)-g(c_i)=(t/c_i)(m-g(c_i))\); its positive part is at most
\((t/a)|e_i-g(c_i)|\), since \(e_i\ge m\).
An outgroup worker uses the larger branch unless \(g(c_i)>m\), with
at most one such exception. All efforts and both branches are bounded.
Thus \(\sum_i |e_i-E(c_i;m)|\) is bounded independently of \(n\) in
either group. Write \(F_n\) for its cost law. The discrepancy between
\(m\) and \(\int E(c;m)\,dF_n(c)\) is \(O(1/n)\). This mean map is
\(t/a\)-Lipschitz. Weak convergence of \(F_n\) and continuity in cost at
\(\eta^*\) now give \(m\to\eta^*\), and efforts converge to
\(e^*(c_i)\) in mean absolute difference.
For each \(\varepsilon>0\), at most one outgroup worker per group can
have \(c_i<c^*-\varepsilon\) for large \(n\), since then \(g(c_i)>m\).
For \(c_i>c^*+\varepsilon\), an ingroup worker must exert at least
\(m\) while \(g(c_i)\) is bounded strictly below \(m\); the bounded
ingroup error sum permits only a bounded number of such workers.
There is no atom at \(c^*\), so the fraction on the wrong side tends
to zero.

\paragraph{Outgroup pairing and task success.}
Changing only outgroup teammates leaves the group means, individual
efforts and the effort each worker expects from any prospective teammate unchanged. Pairs that cease or
begin to be actual teammates in this operation involve two workers
indifferent among outgroup partners. Stability is therefore unchanged.
Pair equal outgroup ranks across the identical lists. Their efforts
then coincide. On the ingroup tasks, the sum of squared effort differences
is bounded by \(2\bar e^2\) across both groups, by the disjoint-interval
bound. With \(w(e)=e+te^2/2\), the identity \(p(e,f)/\kappa=w(e)+w(f)-t(e-f)^2/2\) implies that the loss per task from unequal efforts vanishes. Mean absolute convergence of efforts and boundedness
give limiting success per task \(\int p(e^*(c),e^*(c))\,dF(c)\), exactly
the limit in Proposition~\ref{prop:large-finite-cutoff}.\hfill\(\square\)

\subsection{Blocking pairs at a designer optimum on different finite lists}\label{proof:decentralized-general}

Keep both groups of size \(n\), but now allow different cost lists at every size. Both empirical cost laws converge weakly to the same absolutely continuous \(F\) on the fixed interior cost interval \([a,b]\), as in Proposition~\ref{prop:large-finite-cutoff}. At each optimal matching, all blocking pairs can be covered by a set whose share in each group vanishes along the sequence.
At an optimum, mark an ingroup worker if her teammate exerts less than
the larger group mean, and an outgroup worker if her own effort exceeds
the smaller mean. These marks cover a vanishing fraction of each group.
Indeed both means tend to \(\eta^*\). Except for a vanishing fraction
at costs below \(c^*-\varepsilon\), workers there are ingroup and their
teammates' efforts approach \(g(c_i)>\eta^*\) in mean absolute
difference: the teammate exerts \((c_i e_i-1)/t\), and
\((c_i g(c_i)-1)/t=g(c_i)\). Above \(c^*+\varepsilon\) almost all
workers are outgroup and their efforts are strictly below both means for
large \(n\), by the outgroup best reply. Both inequalities have a uniform
margin away from the cutoff. The limiting mass of the remaining band
vanishes as \(\varepsilon\) decreases.

Among unmarked workers, an ingroup worker will not switch across
groups or prefer a same-group worker currently working outgroup.
Two outgroup workers from the same group cannot both gain, and an
outgroup worker is indifferent among prospects across groups.
The remaining comparison involves two ingroup tasks within a group.
By Proposition~\ref{prop:consecutive-ingroup-ranks}, their pairs occupy
disjoint consecutive ranks in a compatible cost ordering. Efforts in
a task with costs \(x\le y\) lie between \(g(y)\) and \(g(x)\), as
shown in Online Appendix~\ref{proof:decentralized-cutoff}, so both members of the cheaper pair exert at least as
much as either member of the dearer pair. No two unmarked workers
can therefore block. A marked worker can still belong to several
blocking pairs; the argument bounds the size of a set meeting all of
them, not the total number of workers who belong to one.\hfill\(\square\)
\subsection{An asymmetric stable assignment}
\label{oa:asymmetric-stability}

Identical cost lists need not induce symmetric stable assignments
under the bilateral strict comparisons in
Section~\ref{sec:decentralized-matching}. Set \(t=1/2\),
\(\bar e=2\) and \(\kappa=1/8\). In each group, 6 workers have cost
\(3/2\), 20 have cost 3 and 14 have cost 8. In group 1 send the
14 highest-cost workers outgroup; in group 2 send 14 workers of cost 3
outgroup. Pair equal costs ingroup. With \(g,h\) as in the
comparison in Online Appendix~\ref{proof:decentralized-cutoff}, the means satisfy
\[
\begin{aligned}
40\eta_1&=6g(3/2)+20g(3)+14h(8;\eta_2),\\
40\eta_2&=6g(3/2)+6g(3)+14g(8)+14h(3;\eta_1).
\end{aligned}
\]
Their solution satisfies \(\eta_2=15218/38351<g(3)=2/5
<\eta_1=77168/191755\), and \(g(8)<\eta_2\).
Each worker matched outgroup therefore exerts strictly less effort than
she expects from her current teammate. Within a group, two such workers cannot block.
If one proposes to an ingroup worker, the latter currently has a
teammate exerting exactly her own effort: she wants the proposed
partner only if her own effort is even lower, in which case the
proposed partner refuses. Two ingroup workers with equal-effort
teammates cannot both gain within a group either. Across groups an
outgroup worker is indifferent, and every group 1 ingroup worker has
a teammate exerting strictly more than \(\eta_2\). Hence no pair blocks.
In group 2 all highest-cost workers nevertheless work ingroup.

This difference persists with a common absolutely continuous limiting
law. For each integer \(r\), replace the three cost levels in each group
by identical lists of \(3r,10r,7r\) equally spaced midpoints in three
intervals of radius \(\delta\) around the respective costs, with two
workers per midpoint. In group 1 send the whole highest band outgroup;
in group 2 send both copies at the lowest \(7r\) middle-band midpoints.
Pair equal costs ingroup. For all small \(\delta\), the mean equations
are uniform contractions: their sup-norm modulus is at most
\(t/(3/2-\delta)<1\). At the means displayed above, perturbing costs
changes each average by \(O(\delta)\) uniformly in \(r\), since \(g,h\)
are Lipschitz in cost on these intervals. The consistent means therefore
change by \(O(\delta)\) uniformly as well. For fixed sufficiently small
\(\delta>0\), all comparisons with the means used above remain strict;
equal-cost ingroup pairs still exert equal efforts. Each matching is
stable. Both empirical lists converge to the same mixture of the
three uniform laws, but a positive fraction of group 2 works outgroup
at middle costs while all its higher-cost workers stay ingroup.

\section{Analogy partitions and the observation of actions}
\subsection{Equivalence at a fixed matching}
\label{app:ohe-abee}

Fix the finite static game described by $\Gamma^{x,\varphi}$: each player $i\in I$ has characteristic $x_i$, belongs to one of two fixed groups, observes her partner $\varphi(i)$ and both relevant characteristics, and chooses from the common finite action set $A$. A player's payoff depends on the actions and characteristics in her actual pair. Each player chooses once, simultaneously with the others, so her role identifies
her sole decision contingency. An analogy partition $\mathcal C_i$
for player $i$ is a partition of $I\setminus\{i\}$. Write $C_i(j)$
for the class containing $j$. A class-conjecture profile is
$\widehat\beta=(\beta_{i,C})_{i\in I,\ C\in\mathcal C_i}$,
where $\beta_{i,C}\in\Delta(A)$. Player $i$ has the same conjecture $\beta_{i,C}$
about the action of each player in $C$.

\begin{definition}[Analogy-based expectation equilibrium]
\label{def:abee-fixed-matching}
Given the analogy partitions $(\mathcal C_i)_{i\in I}$, an assessment
$(\sigma,\widehat\beta)$ is an analogy-based expectation equilibrium
if the following conditions hold:
\begin{enumerate}
\item \emph{Optimality.} For every $i\in I$ and
$\sigma_i'\in\Delta(A)$,
\[
U_i\bigl(\sigma_i,\beta_{i,C_i(\varphi(i))};
         x_i,x_{\varphi(i)}\bigr)
\geq
U_i\bigl(\sigma_i',\beta_{i,C_i(\varphi(i))};
         x_i,x_{\varphi(i)}\bigr).
\]

\item \emph{Consistency.} For every $i\in I$ and
$C\in\mathcal C_i$,
\[
\beta_{i,C}
=
\frac{1}{|C|}\sum_{j\in C}\sigma_j.
\]
\end{enumerate}
\end{definition}

This specializes the definition in \citet{Jehiel2005} to the
present environment. Every decision contingency is reached with
probability one, giving equal weights within each class.
Strong consistency imposes no additional restriction: any totally
mixed approximation of $\sigma$ has consistent class conjectures
converging to the displayed averages. Conjectures are held fixed
when evaluating deviations, and only the class containing the
actual partner enters a player's payoff calculation.

\paragraph{Group-induced analogy partitions.}
For each player $i$, let $H_i=I\setminus G(i)$ and take
\[
\mathcal C_i
=
\bigl\{\{j\}:j\in G(i)\setminus\{i\}\bigr\}
\cup
\bigl\{H_i:H_i\neq\varnothing\bigr\}.
\]
Thus each other ingroup player forms a singleton class, while all
outgroup players belong to one class, including those whose own
partners are ingroup.

\begin{proposition}
\label{prop:ohe-abee}
Under these group-induced analogy partitions, a mixed-action
profile $\sigma$ is an OHE under $\varphi$ if and only if there
exists a class-conjecture profile $\widehat\beta$ such that
$(\sigma,\widehat\beta)$ is an ABEE. Whenever such an extension
exists, it is unique.
\end{proposition}

\textbf{Proof.}
Consistency uniquely determines every class conjecture. In particular,
\[
\beta_{i,C_i(\varphi(i))}
=
\begin{cases}
\sigma_{\varphi(i)}, & \varphi(i)\in G(i),\\[1mm]
\bar\sigma_{H_i}, & \varphi(i)\in H_i.
\end{cases}
\]
These are exactly the OHE conjectures, so ABEE optimality coincides
with OHE optimality. Consequently, every ABEE yields an OHE profile,
and every OHE extends uniquely to an ABEE by assigning the
consistent averages to all classes. $\blacksquare$

\subsection{Action frequencies by role and by encounter}\label{oa:encounter-feedback}
To compare action frequencies by role and by encounter, allow matching
across roles to follow a distribution
$\mu\in\Delta(\Phi)$. Unlike in the benchmark of Section~\ref{sec:benchmark-matching}, subjects
observe their partner's role even in an outgroup match. Their choices
are conditional only on the role they face. Write
\[
q_{jk}
=
\Pr_{\varphi\sim\mu}\{\varphi(j)=k\},
\qquad
s_j(k)\in\Delta(A)
\quad\text{whenever }q_{jk}>0.
\]
With equal role populations, define the role and group action distributions by
\[
m_j
=
\sum_{k:q_{jk}>0}q_{jk}s_j(k),
\qquad
m_G
=
\frac{1}{|G|}\sum_{j\in G}m_j
\quad\text{for }G\neq\varnothing.
\]

A behavioral contingency is now a pair $(j,k)$, describing role
$j$ acting against role $k$, with frequency $q_{jk}$. For each
observer $i$, role-level ingroup aggregation places all supported
contingencies $(j,k)$ of a given ingroup role $j\neq i$ in one class.
Encounter-specific ingroup aggregation instead makes each such
contingency a singleton. Both specifications place all contingencies
of roles in $H_i=I\setminus G(i)$ in one outgroup class. These partitions are fixed
across matching realizations; class averages use the frequencies
$q_{jk}$.

When $i$ meets $j$, the relevant contingency is $(j,i)$.
The resulting conjectures are
\[
\beta_i^{\mathrm{role}}(j)
=
\begin{cases}
m_j, & j\in G(i),\\
m_{H_i}, & j\in H_i,
\end{cases}
\qquad
\beta_i^{\mathrm{enc}}(j)
=
\begin{cases}
s_j(i), & j\in G(i),\\
m_{H_i}, & j\in H_i.
\end{cases}
\]
The pooled outgroup distribution includes choices in both ingroup and
outgroup encounters.

If $\mu$ assigns probability one to a matching $\varphi$, each role
has only one supported encounter. The two partitions then coincide
on supported contingencies and both reproduce the original OHE
conditions. With random matching, they can yield different
conjectures. Under encounter-specific feedback, best responses in
an ingroup encounter are mutual best responses to the actual
conditional behavior, and hence form a Nash equilibrium of that
pair's game. Role-level feedback need not have this property,
because $m_j$ need not equal $s_j(i)$.

\end{document}